\documentclass[]{interact}
\usepackage[font=small,labelsep=quad,labelfont=bf]{caption}
\usepackage{url}
\usepackage{algorithm, algorithmic}
\usepackage{multirow}
\usepackage{mathrsfs,amsmath,amsfonts,amssymb}
\usepackage{color}
\usepackage{subfigure}
\usepackage{MnSymbol}
\usepackage{hyperref}
\hypersetup{
    colorlinks=true,
    citecolor=black,
    linkcolor=black,
    filecolor=black,
    urlcolor=black,
}
\usepackage{epstopdf}

\usepackage[numbers,sort&compress]{natbib}
\theoremstyle{plain}

\theoremstyle{definition}

\theoremstyle{remark}

\def\hat{\widehat}
\def\tilde{\widetilde}

\def\epsilon{\varepsilon}

\def\*{$\!\!^{^{^{\displaystyle *}}}$}

\def\Rmn#1{\expandafter\uppercase\expandafter{\romannumeral #1}}

\begin{document}

\title{Functional data clustering via information maximization}
\author{
\name{Xinyu Li\textsuperscript{a}, Jianjun Xu\textsuperscript{a*} and Haoyang Cheng\textsuperscript{b}
\thanks{*Corresponding authors. Jianjun Xu . E-mail: xjj1994@mail.ustc.edu.cn (Jianjun Xu)}}
\affil{\textsuperscript{a}International Institute of Finance, School of Management, University of Science and Technology
of China, Hefei, 230026, Anhui, China;\\
 \textsuperscript{b}College of Electrical and Information Engineering, Quzhou University, 324000, Zhejiang, China.}
}

\maketitle

\begin{abstract}
A novel method for clustering functional data is introduced that utilizes information maximization. The method employs unsupervised learning to develop a probabilistic classifier that maximizes the mutual information or squared loss mutual information between data points and their corresponding cluster assignments. A significant advantage of this method is that it involves only continuous optimization of model parameters, which is simpler than discrete optimization of cluster assignments and avoids the drawbacks of generative models. Unlike existing methods, this method does not necessitate the estimation of probability densities of Karhunen-Loeve expansion scores under different clusters and does not require the common eigenfunction assumption. The efficacy of the proposed method is demonstrated through simulation studies and real data analysis. Additionally, the technique permits out-of-sample clustering, and its performance is comparable to that of supervised classifiers.
\end{abstract}
\begin{keywords}
Functional data; Information maximization clustering; Karhunen-Lo\`eve expansion.
\end{keywords}

\section{Introduction}
Advanced collection and storage techniques enable high-frequency data acquisition from a wide variety of research areas, resulting in large amounts of functional data. Examples of such data include stock trading prices in financial markets, near infrared spectroscopy (NIR), birth and death rates in population statistics, temperature and precipitation for consecutive days at national weather stations \cite{levitin2007introduction, ullah2013applications, Wang2018Review,ramsay1991some}. In recent years, clustering analysis has emerged as a popular unsupervised learning method in statistics, with the objective of identifying specific patterns in data that can be easily interpreted. Here are some detailed applications of functional data clustering methods. The first example is the HIV Prevention Trial Network (HPTN) study \cite{zhu2019clustering}. Specifically, the device opening signal generated by electronic medication dispenser device are commonly represented as functional data. Applying functional data clustering methods to adherence monitoring data from such devices can help identify subgroups with distinct medication adherence behavior over the study period. The second example relates to the detection of Nitrogen Oxides (NOx) emission daily curves in the neighborhood of an industrial area. NOx is a major pollutant, and characterizing its behavior is essential for developing appropriate environmental policies. The third example involves the analysis of Martian surface characterization using satellite-based images, which can be converted into spectra with wavelengths ranging from 0.36 to 5.2 microns \cite{bernard2009retrieval}. By clustering these spectroscopic data, we can identify zones composed of similar materials and characterize the composition of the Martian surface. Due to its usefulness, there is no doubt that functional data clustering method has gained tremendous popularity in various fields of sciences, including but not limited to biology, engineering, environmental science, medical science, social science, etc \cite{cheifetz2017modeling,teeraratkul2017shape,gianniou2018clustering,teichgraeber2019clustering}. In this paper, we focus on clustering functional data to identify heterogeneous morphological patterns in the continuous function underlying discrete observations \cite{jacques2014functional}.

Various methods for clustering functional data have been proposed in the academic literature. The first category of methods, known as ``two-stage'' clustering,  which approximates the curves in infinite-dimensional space into a finite-dimensional space of functions spanned by some basis of functions, such as B-spline, wavelet, Fourier, or functional principal component basis, and then applying multivariate clustering algorithms to the basis expansion coefficients \cite{abraham2003unsupervised, james2003clustering, serban2005cats, kayano2010functional}. However, the choice of basis function and the number of basis functions used can significantly affect the clustering results, and their suitability in practical applications is often unknown \cite{2009A}. The second category of methods is distance-based clustering, which utilizes specific distances or dissimilarities between curves along with multivariate non-probabilistic clustering methods. See \cite{peng2008distance,floriello2017sparse,ferraty2006nonparametric,delaigle2019clustering, tokushige2007crisp,meng2018new} for more details. These methods can extract more potential information from the data and improve clustering efficiency in some cases, but it is important to note that there can be substantial differences in the results of clustering methods based on different distances. The third category of methods is adaptive clustering, which is based on probabilistic modeling of basis expansion coefficients. For example, James and Sugar \cite{james2003clustering} proposed a functional data clustering method based on the B-spline basis expansion coefficients. Giacofci et al. \cite{giacofci2013wavelet} proposed the Gaussian model clustering method based on wavelet decomposition (waveclust). Jacques and Preda \cite{jacques2013funclust} proposed a Gaussian mixture clustering model based on Karhunen-Lo{\`e}ve expansion coefficients (funclust). Bouveyron and Jacques \cite{bouveyron2011model} proposed a functional data clustering method based on high-dimensional data clustering (funHDDC). See \cite{chiou2007functional,jiang2012clustering,
bouveyron2015discriminative,chamroukhi2019model,rivera2019robust,heard2006quantitative} for more details. These methods can extract useful information from the data, however, maximizing the joint likelihood may not always lead to the best performance, and clustering results may be incorrect for a breach of the assumption of the generative model. Moreover, some algorithms based on Karhunen-Lo{`e}ve expansion require calculating the eigenfunctions of each cluster, which can lead to a different scale of eigenfunctions due to different means and covariances for each cluster \cite{hall2001functional}. Although the common eigenfunction assumption \cite{zhang2019naive} helps avoid comparing the modes of variation between different bases, verifying this assumption in practice can be challenging.

In the multivariate setting, discriminative clustering techniques have been shown to be capable of directly learning class-posterior probabilities, thereby avoiding the limitations of generative models \cite{kaski2005discriminative}. Barber and  Agakov \cite{barber2005kernelized} and Krause et al. \cite{krause2010discriminative} proposed a discriminative clustering framework based on information maximization. This framework estimates class-posterior probabilities by maximizing an information criterion, and it performs unsupervised classification of samples based on the maximum class-posterior probability criterion. One significant advantage of information maximization clustering is that it transforms the clustering process into a continuous optimization problem, which avoids the discrete optimization of cluster assignments. In addition, information maximization clustering allows for out-of-sample clustering.  Specifically, by inputting the feature variable of a new sample, the corresponding cluster assignment can be directly outputted based on the learned class-posterior probability. Motivated by this method, we extend it to the functional context, which has not been previously explored.

The goal of this paper is to develop a class of functional data clustering techniques that inherits the advantage of multivariate information maximization clustering without estimating the density of Karhunen-Lo{\`e}veexpansion coefficients under different clusters separately. 
Naturally, our method does not require the common eigenfunction assumption. 
Considering that the information criteria are not unique, we propose two functional information maximization clustering methods based on mutual information and square loss mutual information.
In the functional data clustering method based on mutual information, the class-posterior probability is approximated by Karhunen-Lo\`eve expansion and multivariate logistic regression model, and then the mutual information with regularization term is maximized based on L-BFGS quasi-Newton optimization algorithm to obtain the parameters in the multivariate logistic regression model. In the functional data clustering method based on square mutual information, its solution can be computed analytically in a computationally efficient way via kernel eigenvalue decomposition. We provide practical model selection procedures that allow us to objectively optimize tuning parameters in these two methods. Furthermore, this paper discusses corresponding methods for the selection of the number of clusters. Simulation and real data analysis are conducted to exhibit the superiority of the proposed method.

The remainder of the paper is organized as follows. Section \ref{rr1} presents the functional information maximization clustering method based on mutual information and square loss mutual information. Simulation studies and real data analysis are shown in Section \ref{nn1}. Finally, Section \ref{nn2} provides some concluding remarks.

\section{Methodology}\label{rr1}
\subsection{Functional information maximization clustering}
Let $X(t)$ be a functional random variable with values in $L_{2}([0, T]), T>0$. Assume that $X(t)$ is a $L_{2}$-continuous stochastic process. Let $\{x_i(t) \}_{i=1}^{n}$ be an i.i.d sample of size $n$ from the same probability distribution as $X(t)$. We need according to the samples $\{x_i (t) \}_{i = 1}^{n} $, given the corresponding cluster assignments $\left\{y_ {i} \mid y_ {i} \in \{1 \ldots, C \} \right\}_{i = 1}^{n} $, where $C$ is the number of clusters. Given that information criteria are not unique, we use mutual information (MI) and square loss mutual information (SMI) as examples to develop the corresponding functional data clustering method.

Firstly, we present the definitions of MI and SMI. Assume that $Y$ as a categorical variable, $Y=y\in\{1,2,\cdots,C\}$. Let ${Z}$ be a random variable with probability density function $f({z})$. MI and SMI are defined as follows
\begin{equation}\label{aa1}
\mathrm{MI} ({Y}; {Z})=\sum_{y=1}^{C} \int f({z}, {y}) \log \left(\frac{f({z}, {y})}{f({z}) P({y})}\right)d{z},
\end{equation}
\vskip -0.5cm
\begin{equation}\label{aa2}
\mathrm{SMI} ({Y}; {Z})=\frac{1}{2} \sum_{y=1}^{C} \int f({z}) P({y})\left(\frac{f({z},  {y})}{f({z}) P({y})}-1\right)^{2}d{z},
\end{equation}
where $P({y})=P({Y}=y)$, $f(z, y)=f(z|y) P(y)$, $f(z|y)$ denotes the conditional probability density of $Z$ given that $Y=y$. In the concept of divergence, $\mathrm{MI}({Y}; {Z})$ is called Kullback-Leibler divergence \cite{kullback1951information} from $f({z},{y})$ to $f({z})P({y})$, $\mathrm{SMI}({Y};{Z})$ is called Pearson divergence \cite{pearson1900x} from $f({z},{y})$ to $f({z})P({y})$. The Kullback-Leibler divergence and the Pearson divergence both belong to the class of $f$-divergences \cite{csiszar1967information}, and thus they share similar properties. For example, MI and SMI are both non-negative and takes zero if and only if $Z$ and $Y$ are statistically independent. Moreover, MI or SMI is a natural measure of the distance between distributions, this quantity is of particular importance in the field of distributional clustering \cite{banerjee2005clustering}.

Next, it is well established that defining the notion of probability density for a random function is crucial in functional data analysis. A common approach is to employ Karhunen-Lo{`e}ve expansion to approximate the probability density of the random function. In particular, Jacques and Preda \cite{jacques2013funclust} have demonstrated that the probability density of Karhunen-Lo{`e}ve expansion coefficients can serve as a suitable approximation of the probability density of the random function. Specifically, the random function $X(t)$ can be expressed as follows:
\begin{equation}\label{f1}
X(t)=\mu(t)+\sum_{j=1}^{\infty} Z_{j} \psi_{j}(t), \quad j \geq 1,
\end{equation}
where $\mu(t)$ is the mean function of $X(t)$, $Z_{j}=\int_{0}^{T}(X(t)-\mu(t)) \psi_{j}(t) d t, j \geq 1$ are principal components and $\psi_{j}(t)$ 's form an orthonormal system of eigen-functions of the covariance operator of $X(t)$:
\begin{equation}\label{f2}
\int_{0}^{T} \operatorname{Cov}(X(t), X(s)) \psi_{j}(s) d s=\lambda_{j} \psi_{j}(t), \forall t \in[0, T].
\end{equation}
Notice that the principal components are uncorrelated random variables of variance $\lambda_{j}$. Without loss of generality, we assume that $X(t)$ is a random function whose mean value is zero, $\mu(t)=0$. Considering the principal components indexed upon the descending order of the eigenvalues $\left(\lambda_{1} \geq \lambda_{2} \geq \ldots\right)$, let $X^{(q)}(t)$ denotes the approximation of $X(t)$ by truncating (\ref{f1}) at the $q$ first terms, $q \geq 1$,
\begin{equation}\label{a6}
X^{(q)}(t)= \sum_{j=1}^{q} Z_{j} \psi_{j}(t),  \quad q \geq 1.
\end{equation}
Delaigle and Hall \cite{delaigle2010defining} proved that the joint probability density of the random variable $\boldsymbol{Z}^{q}=(Z_1,Z_2,\cdots,Z_q)^\top$ can be used as the approximate probability density of $X(t)$.

Based on Karhunen-Lo{`e}ve expansion, an approximation of the mutual information between the random function $X(t)$ and the categorical variable $Y$ can be obtained as follows:
\begin{equation}\label{aa3}
\begin{aligned}
\mathrm{MI}(Y; \bm{Z}^{q})&:=\int \sum_{y=1}^{C} f(\boldsymbol{z}^{q}(x),  y) \log \frac{f(\boldsymbol{z}^{q}(x),  y)}{f(\boldsymbol{z}^{q}(x)), P(y)}{d}\boldsymbol{z}^{q}(x)\\
&=\int \sum_{y=1}^{C} P(y |\boldsymbol{z}^{q}(x)) f(\boldsymbol{z}^{q}(x))\log P(y | \boldsymbol{z}^{q}(x))d\boldsymbol{z}^{q}(x)\\
&- \int \sum_{y=1}^{C} P(y |\boldsymbol{z}^{q}(x))  f(\boldsymbol{z}^{q}(x)) \log P(y) d\boldsymbol{z}^{q}(x),
\end{aligned}
\end{equation}
where $\bm{z}^{q}(x)=\left(z_{1}(x),  \ldots,  z_{q}(x)\right)^\top$,$\bm{z}^q(x)$ is the first $q$ Karhunen-Lo{\`e}ve expansion coefficients of $x(t)$. $f(\boldsymbol{z}^{q}(x))$ denotes the probability density function of $\bm{Z}^{q}$, $f(\boldsymbol{z}^{q}(x)|y)$ denotes the conditional probability density of $\bm{Z}^{q}$ given that $Y=y$, $f(\boldsymbol{z}^{q}(x), y)=P(y)f(\boldsymbol{z}^{q}(x)|y)$. We use $P(y|\boldsymbol{z}^{q}(x))$ approximate the probability distribution of $Y$ when $x(t)$ is known. Similar to MI, the squared mutual information between the random function $X(t)$ and the categorical variable $Y$ can be approximated as follows:
\begin{equation}\label{aa4}
\begin{aligned}
\mathrm{SMI}(Y; \bm{Z}^{q})&:=\frac{1}{2} \int\sum_{y=1}^{C} f(\boldsymbol{z}^{q}(x)) P({y})\left(\frac{f(\boldsymbol{z}^{q}(x), y)}{f(\boldsymbol{z}^{q}(x)) P({y})}-1\right)^{2} d\boldsymbol{z}^{q}(x)\\
&=\frac{1}{2} \int \sum_{y=1}^{C} f(\boldsymbol{z}^{q}(x)) P({y})\left(\frac{f(\boldsymbol{z}^{q}(x), y)}{f(\boldsymbol{z}^{q}(x)) P({y})}\right)^{2} d{\boldsymbol{z}^{q}(x)}\\
&-\int \sum_{y=1}^{C} f(\boldsymbol{z}^{q}(x)) P({y}) \frac{f(\boldsymbol{z}^{q}(x))}{f(\boldsymbol{z}^{q}(x)) P({y})} d\boldsymbol{z}^{q}(x)+\frac{1}{2} \\
&=\frac{1}{2} \int \sum_{y=1}^{C} P(y \mid \boldsymbol{z}^{q}(x)) f(\boldsymbol{z}^{q}(x)) \frac{P(y \mid \boldsymbol{z}^{q}(x))}{P({y})} d\boldsymbol{z}^{q}(x)-\frac{1}{2}.\\
\end{aligned}
\end{equation}

Then, we regard clustering as an unsupervised classification problem and assume that $Y$ is a latent group variable. Intuitively, on the one hand, $Y$ can be regarded as a kind of encoding of input features, and efficient code $Y$ should be in some way informative about the feature vectors so that the useful information contained in the feature vectors is not lost. Since the information criterion measures the decrease in uncertainty about the feature variable after the information of the class labels is known, it is applicable as a mechanism to generate the code $Y$ \cite{thomas2006elements}.  On the other hand, according to the clustering principles proposed by Chapelle  and Zien \cite{chapelle2005semi}, the decision boundaries should not be located in regions of the input space that are densely populated with data points and the probability distribution of latent group variables should be uniform, which are also referred to as balances class separation and the class of balance. The information maximization clustering satisfies the above two principles, and related work includes \cite{1992Unsupervised,barber2005kernelized,krause2010discriminative,bridle1991unsupervised, grandvalet2004semi}. Based on \eqref{aa3} and \eqref{aa4}, we can estimate $P(y|\boldsymbol{z}^{q}(x))$ by mutual information maximization or square loss mutual information maximization, and these samples are clustered based on the criterion of maximum class-posterior probability
\begin{equation}
{y_i}:=\underset{y\in \left\{1,2,\cdots,C\right\}}{\operatorname{argmax}} P\left(y \mid \boldsymbol{z}^{q}(x_i)\right),
\end{equation}
where $\bm{z}^q(x_i)$ is the first $q$ Karhunen-lo{\`e}ve expansion coefficients of $x_i(t)$. In subsections \ref{rr2} and \ref{rr3}, the estimation of $P(y|\boldsymbol{z}^{q}(x))$, $P(y)$ and truncation number $q$ are introduced. In addition, it is worth noting that cluster number $C$ is assumed to be known during the development of the proposed method, but $C$ is usually unknown in practice. In subsection \ref{rr4}, we discuss various methods for determination of the number of clusters.

\subsection{Functional data clustering based on MI maximization}\label{rr2}
Assuming that the class-posterior probability $P(y|\boldsymbol{z}^{q}(x))$ is a parametric model with the parameter $\boldsymbol{W}$, the corresponding parameter model is denoted as $P\left(y \mid \boldsymbol{z}^{q}(x);\bm{W}\right)$. In this case, an approximation of $P(y)$ can be obtained as follows:
\begin{equation}
\hat{P}(y)= \frac{1}{n} \sum_{i=1}^{n} P(y |\boldsymbol{z}^{q}(x_i); \bm{W}).
\end{equation}
Due to $\bm{W}$ being unknown, the aforementioned equation cannot be referred to as an ``estimate'' and is instead considered an ``approximation''. The approximation of \eqref{aa3} can be obtained as follows:
\begin{equation}\label{vvvv2}
\begin{aligned}
\mathrm{MI}_{\bm{W}}(Y; \bm{Z}^{q}):=& \frac{1}{n} \sum_{y=1}^{C}\sum_{i=1}^{n} 
P\left(y \mid \bm{z}^{q}(x_i) ; \bm{W}\right) \log P\left(y \mid \bm{z}^{q}(x_i) ; \bm{W}\right) \\
&-\frac{1}{n}\sum_{y=1}^{C}\sum_{i=1}^{n} P\left(y \mid \boldsymbol{z}^{q}(x_i) ; \bm{W}\right) \log \left(\frac{1}{n} \sum_{i=1}^{n}
P\left(y \mid \bm{z}^{q}(x_i) ; \bm{W}\right)\right).
\end{aligned}
\end{equation}
Bridle \cite{bridle1991unsupervised} and Krause et al. \cite{krause2010discriminative} have noted that classifiers trained with the objective function MI tend to fragment the data into a large number of categories. To address this issue, a regularizing term $R(\lambda ;\boldsymbol{W})$ is introduced. The specific form of this term depends on the choice of $P\left(y \mid \bm{z}^{q}(x) ; \bm{W}\right)$ and penalizes conditional models with complex decision boundaries in order to yield sensible clustering solutions. The following regularization mutual information maximization problem is solved to estimate $\bm{W}$:
\begin{equation}\label{mmmm1}
\begin{aligned}
& \underset{\boldsymbol{W}}{\operatorname{argmax}}{~\mathrm{MI}_{\bm{W}}(Y; \bm{Z}^{q})-R(\lambda ;\boldsymbol{W})}\\
&= \underset{\boldsymbol{W}}{\operatorname{argmax}} \frac{1}{n} \sum_{i=1}^{n} \sum_{y=1}^{C} P\left(y \mid \bm{z}^{q}(x_i) ; \bm{W}\right)
\log P\left(y \mid \bm{z}^{q}(x_i) ; \bm{W}\right)\\
&-\frac{1}{n}\sum_{y=1}^{C}\sum_{i=1}^{n} P\left(y \mid \boldsymbol{z}^{q}(x_i) ; \boldsymbol{W}\right) \log \left(\frac{1}{n} \sum_{i=1}^{n} P\left(y \mid \boldsymbol{z}^{q}(x_i) ; \boldsymbol{W}\right)\right)- R(\lambda ;\bm{W}),
\end{aligned}
\end{equation}
where $\lambda$ is a regularization parameter. The framework is flexible in the choice of $P\left(y \mid \bm{z}^{q}(x) ; \bm{W}\right)$ and $R(\lambda ;\boldsymbol{W})$. As an example instantiation, we here choose multiclass logistic regression as the conditional model 
\begin{equation}\label{vvvv3}
	P\left(Y=y \mid \boldsymbol{z}^{q}(x) ; \boldsymbol{W}\right) \propto \exp \left(\boldsymbol{\alpha}^{\top}_y\boldsymbol{z}^{q}(x)+b_{y}\right),
\end{equation}
and the regularization term is the squared $L_{2}$ norm of the weight vectors
\begin{equation}
	R(\lambda, \boldsymbol{\alpha}):= \lambda\sum_{y=1}^{C}\bm{\alpha}^{\top}_y\bm{\alpha}_y,
\end{equation}
where each weight vector $\boldsymbol{\alpha}_y \in \mathbb{R}^{q}$ is $q$-dimensional with components $\alpha_{c q}$, $\bm{\alpha}_y=({\alpha_{y1}}, {\alpha_{y2}}, \cdots, {\alpha_{yq}})^{\top}$. The set of parameters $\boldsymbol{W}=\left\{\boldsymbol{\alpha}_1, \boldsymbol{\alpha}_2, \cdots, \boldsymbol{\alpha}_C, b_1, \cdots, b_C\right\}$ consists of weight vectors $\boldsymbol{\alpha}_{y}$ and bias values $b_{y}$ for each cluster $y$, $\boldsymbol{\alpha}=(\boldsymbol{\alpha}_1, \cdots, \boldsymbol{\alpha}_C)$. Note that the bias terms $\{b_c\}_{c=1}^{C}$ are not penalized. The parameters $\boldsymbol{W}$ can be estimated by using L-BFGS quasi-Newton optimization algorithm \cite{liu1989limited}. Specifically, the partial derivatives of ${\alpha}_{y j}$ and ${b}_{y}$ are as follows:
\begin{equation}\label{vvvv4}
\begin{aligned}
&\frac{\partial \mathrm{MI}_{\bm{W}}(Y; \bm{Z}^{q})-R(\lambda ;\bm{\alpha})}{\partial  {\alpha_{y j}} }=\frac{1}{n} \sum_{i=1}^{n}
 \bm{z}_j(x_i)p_{y i}\left(\log \frac{p_{y i}}{{p}_{y}}-\sum_{y=1}^{C} p_{y i} \log \frac{p_{y i}}{{p}_{y}}\right)-2 \lambda {\alpha_{y j}}, \\
&\frac{\partial \mathrm{MI}_{\bm{W}}(Y; \bm{Z}^{q})-R(\lambda ;\bm{\alpha})}{\partial b_{y}}=\frac{1}{n} \sum_{i=1}^{n} p_{y i}\left(\log \frac{p_{y i}}{{p}_{y}}-\sum_{y=1}^{C} p_{y i} \log \frac{p_{y i}}{{p}_{y}}\right), j=1,2,\cdots,q,\\
\end{aligned}
\end{equation}
where $p_{y i} \equiv P(Y=y |\boldsymbol{z}^{q}(x_i), \boldsymbol{W})$, ${p}_{y} \equiv \hat{P}(Y=y,\boldsymbol{W})$. Finally, the assignment of each sample to a cluster is determined based on the class-posterior probability
\begin{equation}\label{vvvv5}
\widehat{y}_{i}=\underset{y\in \left\{1,2,\cdots,C\right\}}{\operatorname{argmax}} P\left(y \mid \boldsymbol{z}^{q}(x_i) ; \widehat{\boldsymbol{W}}\right),
\end{equation}
where $\widehat{\boldsymbol{W}}$ is the estimator obtained by L-BFGS quasi-Newton optimization algorithm. To initialize the parameter $\boldsymbol{W}$, we adopt the initialization method proposed by Krause et al. \cite{krause2010discriminative}. Additionally, we can utilize \eqref{vvvv5} to predict the cluster assignment of a new sample $x(t)$. We call the above method ``Functional Mutual Information maximizing clustering"(FMIclust).

The effectiveness of FMIclust relies on the selection of the truncation number $q$ and the regularization parameter $\lambda$. The estimation of the truncation number $q$ is an open problem with no unique technique to use. Much of the existing literature, such as \cite{jacques2013funclust,bouveyron2011model,BOUVEYRON2007502} suggested that $q$ can be chosen subjectively. Specifically, the truncation number $q$ is estimated through the scree-test of Cattell \cite{1966The} which looks for a break in the eigenvalues scree. The selected $q$ is the one for which the subsequent eigenvalues differences are smaller than a threshold. The threshold can be provided by the user or selected using BIC \cite{1978Estimating}. In the paper, we recommend using BIC which yields satisfactory results. We then consider the selection of $\lambda$ when $q$ is known. Noting that the labeled samples have already obtained in the model selection stage and thus supervised estimation of MI is possible. For supervised MI estimation, there exists a more powerful supervised MI estimator called maximum-likelihood MI (MLMI), which was proved to achieve the optimal non-parametric convergence rate \cite{suzuki2008approximating}. To select $\lambda$, we aim to maximize $\mathrm{MLMI}(\hat{\bm{Y}}; \boldsymbol{Z}^{q})$. The detailed calculation procedure of MLMI can be found in \cite{suzuki2008approximating}. Finally, the FMIclust algorithm can be summarized as follows:
\begin{algorithm}
	\renewcommand{\algorithmicrequire}{\textbf{Input:}}
	\renewcommand{\algorithmicensure}{\textbf{Output:}}
	\caption{FMIclust}
	
	\begin{algorithmic}[1]
		\REQUIRE $\{x_i(t),  t \in \mathcal{I} \}_{i=1}^{n}$ and $C$ ; $\lambda \in \Lambda$ and $\Lambda$ is the set of regularization parameter.
		\ENSURE $\left\{y_{i} \mid y_{i} \in\{1,  \ldots,  C\}\right\}_{i=1}^{n}$.
\STATE  calculating $\{\bm{z}^{q}(x_i) \}_{i=1}^{n}$ based on Karhunen-Lo{\`e}ve expansion, where $q$ is selected based on the scree-test of Cattell.
		\STATE  initialize parameters $\bm{W}$;
    		   \FORALL{$\lambda \in \Lambda$}
    		\STATE according to \eqref{mmmm1}-\eqref{vvvv4}, The L-BFGS quasi-Newton optimization algorithm is used to obtain the optimal parameter $\widehat{\bm{W}}$;
    		\STATE  $\hat{\bm{Y}} = \left\{\hat{y}_{i} \right\}_{i=1}^{n}$ from (\ref{vvvv5});
    		\STATE  using the labeled samples $\left\{\left(\bm{z}^{q}(x_i),  \hat{y}_{i}\right)\right\}^{n}_{i=1}$ calculate $\mathrm{MLMI}(\hat{\bm{Y}};\bm{Z}^{q},\lambda)$;
    		\ENDFOR
		\STATE \textbf{return} $\hat{\bm{Y}}$ corresponding to the maximum $\mathrm{MLMI}(\bm{Y};\bm{Z}^{q},\lambda)$
	\end{algorithmic}
\end{algorithm}

\subsection{Functional data clustering based on SMI maximization}\label{rr3}
Assuming that the class-prior probability $P(y)$ is set to a user-specified value $\pi_{y}$ for $y=1, \ldots, C$, where $\pi_{y}>0$ and $\sum_{y=1}^{C} \pi_{y}=1$. Without loss of generality, we assume that $\left\{\pi_{y}\right\}_{y=1}^{C}$ are sorted in the ascending order, $\pi_{1} \geq \cdots \geq \pi_{C}$. If class-prior distribution is unknown, we may adopt the uniform distribution:
$\pi_y = 1/C$. See \cite{liu1989limited,shi2000normalized,xu2004maximum,niu2013maximum,zelnik2004self} for more details. According to the work of \cite{sugiyama2014information}, the class-posterior probability $P(y \mid \bm{z}^{q}(x))$ can be approximated using kernel expansion as follows:
\begin{equation}
P(y \mid \bm{z}^{q}(x);\bm{\beta}_y):=\sum_{i=1}^{n} \beta_{yi} \mathcal{K}\left(\bm{z}^{q}(x),  \bm{z}^{q}(x_i)\right),
\end{equation}
where $\bm{\beta}_y=(\beta_{y1}, \cdots, \beta_{yn})^\top$ is a parameter vector, and $\mathcal{K}\left(\bm{z}^{q}(x),  \bm{z}^{q}(x_i)\right)$ denotes a kernel function. We then use a sparse variant of the local-scaling kernel \cite{zelnik2004self}:
\begin{align}
&\mathcal{K}\left(\bm{z}^{q}(x_i), \bm{z}^{q}(x_{i^{'}})\right)\notag\\
&=\left\{\begin{array}{lc}
\exp \left(-\frac{\left\|\bm{z}^{q}(x_i)-\bm{z}^{q}(x_{i^{'}})\right\|^{2}}{2 \sigma_{i} \sigma_{{i^{'}}}}\right),
& \bm{z}^{q}(x_i) \in \mathcal{N}_{v}\left(\bm{z}^{q}(x_{i^{'}})\right) \quad \text{or} \quad  \bm{z}^{q}(x_{i^{'}}) \in \mathcal{N}_{v}\left(\bm{z}^{q}(x_i)\right), \\
0, & \text {else},
\end{array}\right.
\end{align}
where $i,i^{'}=1,2,3,\cdots,n$. $\mathcal{N}_{v}\left(\bm{z}^{q}(x_{i})\right)$ denotes the set of $v$ nearest neighbors for $\bm{z}^{q}(x_i)$ ( $v$ is the kernel parameter), $\bm{z}^{q}(x_i)^{(v)}$ is the $v$-th nearest neighbor of $\bm{z}^{q}(x_i)$, $\sigma_{i}$ is a local scaling value defined as $\sigma_{i}=\left\|\bm{z}^{q}(x_i)-\bm{z}^{q}(x_i)^{(v)}\right\|_{\mathbb{R}^q}$. By using empirical approximation, the sample form of the mean square loss mutual information maximization is as follows.
\begin{equation}\label{qwe1}
\underset{\bm{\beta}}{\operatorname{argmax}}{~\widehat{\mathrm{SMI}}(Y; \bm{Z}^{q})}:=
\underset{\bm{\beta}}{\operatorname{argmax}}{~\frac{1}{2n} \sum_{y=1}^{C} \frac{1}{\pi_y}\bm{\beta}_{y}^\top \bm{\mathcal{K}}^\top \bm{\mathcal{K}} \bm{\beta}_{y}-\frac{1}{2}},
\end{equation}
where $\bm{\beta}=\{\bm{\beta}_y\}_{y=1}^C$, the $(i,{i^{'}})$-th element of matrix $\bm{\mathcal{K}}$ is $\bm{\mathcal{K}}_{i,{i^{'}}}= \mathcal{K}\left(\boldsymbol{z}^{q}(x_i), \bm{z}^{q}(x_{i^{'}})\right)$. A natural optimization criterion would be to impose non-negativity and normalization constraints on the parameter $\bm{\beta}$. In order to obtain the global optimal solution analytically even though the optimization problem is still non-convex, we use the unit-norm constraint $\|\bm{\beta}_y\|_{\mathbb{R}^n}=1$ on the parameter, where $\|\cdot\|_{\mathbb{R}^n}$ is the $L_1$ norm of $\bm{\beta}_y$, $\|\bm{\beta}_y\|_{\mathbb{R}^n}=\sqrt{{\beta}^{2}_{y1}+\cdots+{\beta}^{2}_{yn}}$. Futhermore, (\ref{qwe1}) can be described as maximizing $\bm{\beta}_{y}^\top \bm{\mathcal{K}}^\top\bm{\mathcal{K}} \bm{\beta}_{y}$ for each cluster $y$ under the condition of $\left\|\bm{\beta}_{y}\right\|_{\mathbb{R}^n}=1$, so the maximizer is given by the normalized principal eigenvector of $\bm{\mathcal{K}}$. 
To avoid all the solutions $\bm{\beta}=\left\{\beta_y \right\}_{y=1}^{C}$ to be reduced to the same principal
eigenvector, we impose their mutual orthogonality: $\bm{\beta}_y^{\top}\bm{\beta}_{y^{\prime}}=0$ for $y \neq y^{\prime}$. Then the solutions are given by the normalized eigenvectors $\boldsymbol{\eta}_{1}, \ldots, \boldsymbol{\eta}_{c}$ associated with the eigenvalues $\lambda_{1} \geq \cdots \geq \lambda_{n} \geq 0$ of $\bm{\mathcal{K}}$. Since the sign of $\boldsymbol{\eta}_{y}$ is arbitrary, we set the sign as
$$
\widetilde{\boldsymbol{\eta}}_{y}=\boldsymbol{\eta}_{y} \times \operatorname{sign}\left(\boldsymbol{\eta}_{y}^{\top}\mathbf{1}_{n}\right),
$$
where $\mathbf{1}_{n}$ is n-dimensional column vectors with all ones. $\mathrm{sign}(\cdot)$ is sign function. On the other hand, considering class-prior probability $P(y)$ can be written as:
$$
P(y)=\int P(y \mid \bm{z}^{q}(x)) f(\bm{z}^{q}(x))d\bm{z}^{q}(x) \approx \frac{1}{n} \sum_{i=1}^{n}
P\left(y \mid \bm{z}^{q}(x_i); \bm{\beta}_y\right)=\frac{1}{n} \bm{\beta}_{y}^{\top} \bm{\mathcal{K}} \mathbf{1}_{n},
$$
and the class-prior probability $P(y)$ was set to $\pi_{y}$ for $y=1, \ldots, c$, we have the following normalization condition:
$$
\frac{1}{n} \bm{\beta}_{y}^{\top} \bm{\mathcal{K}} \mathbf{1}_{n}=\pi_{y}.
$$
Additionally, probability estimates should be non-negative, which can be achieved by
rounding up negative outputs to zero. The cluster assignment corresponding to each sample is determined based on the class-prior probability $P(y |\bm{z}^{q}(x))$.
\begin{equation}\label{v11}
\widehat{y}_{i}=\underset{y \in \{1, 2, \cdots, C\}}{\operatorname{argmax}} \frac{\left[\max \left(\mathbf{0}_{n},  \bm{\mathcal{K}} \widetilde{\boldsymbol{\eta}}_{y}\right)\right]_{i}}{(n\pi_y)^{-1} \max \left(\mathbf{0}_{n},  \boldsymbol{\mathcal{K}} \widetilde{\boldsymbol{\eta}}_{y}\right)^{\top} \mathbf{1}_{n}}=\underset{y \in \{1, 2, \cdots, C\}}{\operatorname{argmax}} \frac{\pi_y \left[\max \left(\mathbf{0}_{n},  \widetilde{\boldsymbol{\eta}}_{y}\right)\right]_{i}}{\max \left(\mathbf{0}_{n},  \widetilde{\boldsymbol{\eta}}_{y}\right)^{\top} \mathbf{1}_{n}},
\end{equation}
where $\mathbf{0}_{n}$ denotes the $n$-dimensional vector with all zeros, the max operation for vectors is applied in the element-wise manner, and $[\cdot]_{i}$ denotes the $i$-th element of a vector. For out-of-sample prediction, cluster assignment $y^{new}$ for new sample ${x}^{new}_{i}$ may be obtained as
\begin{equation}\label{S22}
y^{new}:=\underset{y \in \{1, 2, \cdots, C\}}{\operatorname{argmax}} \frac{\pi_{y} \max \left(0,  \sum_{i=1}^{n} \mathcal{K}\left({\bm{z}}^{q}(x),  \bm{z}^{q}({x}^{new}_{i})\right)\left[\tilde{\boldsymbol{\eta}}_{y}\right]_{i}\right)}{\lambda_{y} \max \left(\mathbf{0}_{n},  \widetilde{\boldsymbol{\eta}}_{y}\right)^{\top} \mathbf{1}_{n}}.
\end{equation}
We call the above method ``Functional Square loss Mutual Information maximizing clustering''(FSMIclust). 

The choice method of truncation number $q$ is consistent with FMIclust. Additionally, since FSMIclust was developed in the framework of SMI maximization, it would be natural to determine the kernel parameter $v$ by maximizing SMI. A direct approach is to use the SMI estimator ($\widehat{\text{SMI}}$) in this paper for kernel parameter choice. However, this direct approach is not favorable since $\widehat{\text{SMI}}$ is an unsupervised SMI estimator. Fortunately, in the model selection stage, we have already obtained labeled samples and thus supervised estimation of SMI is possible. For supervised SMI estimation, a nonparametric estimator called least square loss mutual information (LSMI), which was proved to achieve the optimal convergence rate \cite{suzuki2009mutual}. In principle, it is possible to use an arbitrary clustering algorithm in the first step and then evaluate its validity by LSMI in the second stage. To select the kernel parameter $v$, we aim to maximize $\mathrm{LSMI}(\hat{\bm{Y}}; \boldsymbol{Z}^{q})$. The detailed calculation procedure of LSMI can be found in \cite{suzuki2009mutual}. Finally, the FSMIclust algorithm can be summarized as follows:
\begin{algorithm}
	\renewcommand{\algorithmicrequire}{\textbf{Input:}}
	\renewcommand{\algorithmicensure}{\textbf{Output:}}
	\caption{FSMIclust}
	\label{alg:1}
	\begin{algorithmic}[1]
		\REQUIRE $\{x_i(t),  t \in \mathcal{I} \}_{i=1}^{n}$ and $C$; $v \in \mathcal{V}$ and $\mathcal{V}$ is a set of the kernel parameter.
		\ENSURE $\left\{y_{i} \mid y_{i} \in\{1,  \ldots,  C\}\right\}_{i=1}^{n}$.
		\STATE  calculating $\{\bm{z}^{q}(x_i) \}_{i=1}^{n}$ based on Karhunen-Lo{\`e}ve expansion, where $q$ is selected based on the scree-test of Cattell;
            \FORALL{$v \in \mathcal{V}$}
    		\STATE calculating kernel matrix $\boldsymbol{\mathcal{K}}$;
    		\STATE calculating the unit orthogonalized eigenvectors $\boldsymbol{\eta}_y$ corresponding to the first C largest eigenvalues of $\bm{\mathcal{K}}$ , $y=1, 2, \cdots, C$;
    		\STATE $\widetilde{\boldsymbol{\eta}}_{y} \longleftarrow \boldsymbol{\eta}_y \times
                   \operatorname{sign}\left(\bm{\eta}_{y}^{\top} \mathbf{1}_{n}\right),~y=1,  2, \ldots, C$;
    		\STATE $\hat{\bm{Y}} =\left\{\hat{y}_{i}\right\}_{i=1}^{n}$ from \eqref{v11};
    		\STATE  using the labeled samples $\left\{\left(\bm{z}^{q}(x_i),  \hat{y}_{i}\right)\right\}^{n}_{i=1}$ calculate $\mathrm{LSMI}(\hat{\bm{Y}};\bm{Z}^{q},v)$;
                \ENDFOR
		\STATE \textbf{return} $\hat{\bm{Y}}$ corresponding to the maximum $\mathrm{LSMI}(\hat{\bm{Y}};\bm{Z}^{q},v)$.
	\end{algorithmic}
\end{algorithm}

\subsection{Estimation of the cluster numbers C}\label{rr4}
In the estimation process, we assume that the number of clusters $C$ is known, but $C$ is usually unknown in practice, so it is necessary to estimate the number of clusters. For FMIclust, Krause et al. \cite{krause2010discriminative} show that the regularization term $R(\lambda, \boldsymbol{\alpha})$ is composed of the sum of penalty terms $ \sum_{i, j=1}^{n} \boldsymbol{\alpha}^{\top}_y\boldsymbol{\alpha}_y$ associated with each cluster. Setting the derivatives \eqref{vvvv4} equal to zero yields the following condition at stationary points of \eqref{mmmm1},we have
$$
\boldsymbol{\alpha}_y = \sum_{i}^{n} \frac{1}{2 \lambda n} p_{y i}\left(\log \frac{p_{y i}}{{p}_{y}}-\sum_{y}^{C} p_{y i} \log \frac{p_{y i}}{{p}_{y}}\right)\boldsymbol{z}^{q}(x_i).
$$
It is instructive to observe the limiting behavior of the regularization term $\boldsymbol{\alpha}^{\top}_y\boldsymbol{\alpha}_y$ when datapoints are not assigned to cluster $y$; that is, when $\hat{p}_{y}=$ $\frac{1}{N} \sum_{i} p_{y i} \rightarrow 0$. This implies that $p_{y i} \rightarrow 0$ for all $i$, and therefore $\boldsymbol{\alpha}^{\top}_y\boldsymbol{\alpha}_y \rightarrow 0$. This means that the regularizing function does not penalize unpopulated clusters. Based on this, we can initialize a large number of clusters, then based on corresponding penalty terms of all clusters sorted in descending order, and we can use the scree-test of Cattell and choose the ``elbow'' point as the number of clusters.

Regarding FSMIclust, it is worth noting that the clustering solution is obtained by solving the eigenvector problem of the kernel matrix. Consequently, the modified Bayesian Information Criterion (MBIC) can be employed to determine the optimal number of clusters, as described in \cite{wang2016functional}. Specifically, based on eigenvalue estimators sorted in descending order $\left\{\hat{\lambda}_i; i,2,\cdots,n \right\}$, the MBIC criterion is defined as follows:
$$
\operatorname{MBIC}(C)=\sqrt{n} \frac{\sum_{i=1}^{C}\widehat{\lambda}_i}{\sum_{i=1}^{n}\widehat{\lambda}_i}-\frac{2 \log (n) C}{n},
$$
where $C$ represents the number of clusters to be estimated. The optimal number of clusters is obtained by selecting the value of $C$ that maximizes the MBIC criterion over the interval $1 \leq C \leq n-1$.

\section{Data analysis}\label{nn1}
In this section, we compare FMIclust and FSMIclust with existing functional clustering methods. Simulation studies are conducted to provide empirical performance of the proposed method and compare it with some existing methods including distclust \cite{peng2008distance}, waveclust \cite{giacofci2013wavelet}, funclust \cite{jacques2013funclust}, funHDDC \cite{bouveyron2011model}, iterSubspace \cite{chiou2007functional}, fscm \cite{jiang2012clustering}, K-means clustering based on B-spline basis expansion (B-Kmeans), K-means clustering based on the expansion of functional principal component basis (FPCA-Kmeans) \cite{hartigan1979algorithm}, Gaussian mixture model based on B-spline basis expansion (B-GMM), Gaussian mixture model based on functional principal component basis expansion (FPCA-GMM) \cite{celeux1995gaussian}. Real data analysis is conducted in four application scenarios, including comparing male and female growth curves, distinguishing myocardial infarction and normal ECG curves, dividing the urban climatic zone in Canada, and analyzing the current situation of COVID-19 epidemic risk in China. The performance of clustering is evaluated using Purity Function (PF) and Adjusted Rand Index (ARI) measurements. A higher PF and ARI indicate better clustering results.

To implement these methods, the B-spline basis expansion and functional principal component analysis can be carried out using the R package  ``fda'' (\url{https://cran.r-project.org/web/packages/fda/index.html}). Kmeans and GMM can be implemented using the R program ``kmeans'' (\url{https://www.rdocumentation.org/packages/stats/versions/3.6.2/topics/kmeans}) and ``Mclust'' (\url{https://cran.r-project.org/web/packages/mclust/}). Distclust, iterSubspace, waveclust, fscm, funclust, and funHDDC can be implemented using the R package "funcy" (\url{https://rdrr.io/cran/funcy/}), with the tuning parameters of comparison methods determined by their respective references, which can be easily obtained through the R package ``funcy''. For the proposed method, Karhunen-Lo{\`e}ve expansion can be implemented using the MATLAB program called ``PACE'' (\url{https://github.com/functionaldata/PACE_matlab/blob/master/release2.17/PACE)}. L-bfgs quasi-Newton optimization algorithm can be implemented using the MATLAB program called  ``minFunc''(\url{http://www.cs.ubc.ca/schmidtm/Software/minFunc.html}). For FSMIclust, the kernel matrix decomposition can be implemented using the MATLAB program called ``eigs'' and The choice method of kernel parameter $v$ can be implemented using the MATLAB program ``SMIC" (\url{http://www.ms.k.u-tokyo.ac.jp/sugi/software.html#SMIC}), the prior probability $\pi_y$ is estimated using the uninformative prior probability. LSMI can be implemented using the MATLAB program called ``LSMI'' (\url{http://www.ms.k.u-tokyo.ac.jp/sugi/software.html#LSMI}).

\subsection{Simulation study}
First, the design for the Simulation 1 comes from \cite{jacques2013funclust}, and the generation process is shown as follows:
\begin{equation}
X^{c}_i(t)=U_{c, 1} h_{1}(t)+U_{c, 2} h_{2}(t)+\epsilon,  \quad t \in[1, 21]
\end{equation}
where $c=1$ and $2$. $U_{1,1}$, $U_{1,2}$, $U_{2,1}$ and $U_{2,2}$ are all random variables following the Gaussian distribution, $\mathbb{E}\left[U_{1, 1}\right]=\mathbb{E}\left[U_{1, 2}\right]=0$,  $\operatorname{Var}\left(U_{1, 1}\right)=\operatorname{Var}\left(U_{1, 2}\right)=1/12$,  $\mathbb{E}\left[U_{2, 1}\right]=\mathbb{E}\left[U_{2, 2}\right]=0.05$,  $\operatorname{Var}\left(U_{2, 1}\right)=\operatorname{Var}\left(U_{2, 2}\right)=1/6$. $\epsilon$ is a white noise and independent of $\{U_{i,j};~i,j=1,2\}$, where $\operatorname{Var}\left(\epsilon\right)=1/12$.
$h_{1}(t)=6-|t-7|$,  $h_{2}(t)=6-|t-15|$.
The 100 observation points of each curve are distributed at equal intervals [1,21]. Assume that the number of samples for each cluster is 100, and the experiment is repeated 100 times.

Second, the design for the Simulation 2 comes from \cite{zhong2021cluster}, and the generation process is shown as follows:
\begin{equation}
X_{y}(t)=\sum_{k=1}^{2} \xi_{yk}\phi_{yk}(t)+\varepsilon_{yt},
\end{equation}
where $y=1, 2, 3$, $\phi_{11}(t)=\sqrt{2} \cos ( \pi t)$,
$\phi_{12}(t)=\sqrt{2} \sin (\pi t)$, $\phi_{21}(t)=\sqrt{2} \cos (2 \pi t)$,  $\phi_{22}(t)=\sqrt{2} \sin (\pi t)$,  $\phi_{31}(t)=\sqrt{2} \cos (2 \pi t)$, $\phi_{32}(t)=\sqrt{2} \cos (\pi t)$.  $\xi_{yk} \sim N\left(\theta_{yk},  \lambda_{yk}\right)$,
where $\theta_{1k}=0.3k$,  $\lambda_{11}=1/6$,  $\lambda_{12}=1/12$,  $\theta_{2k}=0.5k$,  $\lambda_{21}=1/3$,  $\lambda_{22}=1/6$,  $\theta_{3k}=0.1k$,  $\lambda_{31}=1/3$,  $\lambda_{32}=1/12$.
Random error $\varepsilon_{yt} \sim N\left(0,  \sigma_{y}^{2}\right)$, where $\sigma_{1}^{2}=0.1$,  $\sigma_{2}^{2}=0.15$,  $\sigma_{3}^{2}=0.2$. $\{\varepsilon_{yt},~ y=1, 2, 3\}$ are independent of each other. The 100 observation points are equally spaced in the interval [0,1]. Assume that the number of samples for each cluster is 100, and the experiment is repeated 100 times.

\begin{table}[ht]
\begin{center}
\caption{PF and ARI unde different methods (Simulation 1)}
\vskip -0.5cm
\footnotesize
\begin{tabular}{cccccccccc}\label{mv1}
\toprule
\multirow{1}{*}{method} & \multicolumn{1}{c}{PF} & \multicolumn{1}{c}{ARI} &\multirow{1}{*}{method} & \multicolumn{1}{c}{PF} & \multicolumn{1}{c}{ARI}\\\midrule
FMIclust            & 0.9140(0.0241)   & 0.8495(0.0362)   &waveclust            & 0.6730(0.0773)    & 0.5260(0.0732)     \\
FSMIclust           & 0.9340(0.0352)   & 0.8645(0.0262)   & fscm                & 0.8406(0.0463)    & 0.7534(0.0467)    \\
funHDDC             & 0.8693(0.0532)   & 0.8058(0.0437)   & B-Kmeans            & 0.8632(0.0684)    & 0.8039(0.0748)    \\
funclust            & 0.8837(0.0635)   & 0.8286(0.0826)   &B-GMM                & 0.7880(0.0484)    & 0.6593(0.0577)     \\
distclust           & 0.8619(0.0421)   & 0.7945(0.0524)   &FPCA-Kmeans          & 0.8097(0.0672)    & 0.7287(0.0621)     \\
iterSubspace        & 0.8440(0.0374)   & 0.7657(0.0216)   &FPCA-GMM             & 0.8564(0.0634)    & 0.7743(0.0579)     \\\bottomrule
\end{tabular}
\end{center}
\vskip -0.5cm
\end{table}

For each clustering method, we report the mean and standard deviation of the partitioning fidelity (PF) and adjusted Rand index (ARI) across 100 simulations in Tables \ref{mv1} and \ref{mv2}. Simulation 1 assumes that the eigenfunction of different clusters is identical, while Simulation 2 assumes that the eigenfunction of different clusters is distinct. Based on the clustering evaluation results, it is evident that the proposed method outperforms the other clustering approaches in both simulation scenarios.

\begin{table}[ht]
	\begin{center}
		\caption{PF and ARI under different methods (Simulation 2)}
		\vskip -0.5cm
		\footnotesize
		\begin{tabular}{cccccccccc}\label{mv2}
			\toprule
			\multirow{1}{*}{method} & \multicolumn{1}{c}{PF} & \multicolumn{1}{c}{ARI} &\multirow{1}{*}{method} & \multicolumn{1}{c}{PF} & \multicolumn{1}{c}{ARI}\\\midrule
			FMIclust            & 0.8895(0.0267)     & 0.7712(0.0259)   &waveclust           & 0.8644(0.1373)    & 0.7586(0.1493)    \\
			FSMIclust           & 0.8937(0.0226)     & 0.7939(0.0326)   &fscm                & 0.8409(0.0289)    & 0.7502(0.0312)     \\
			funHDDC             & 0.7872(0.1213)     & 0.6771(0.1325)   &B-Kmeans            & 0.7347(0.0668)    & 0.6149(0.0765)     \\
			funclust            & 0.6321(0.0833)     & 0.4287(0.1118)   &B-GMM               & 0.6775(0.0853)    & 0.5790(0.0867)     \\
			distclust           & 0.7924(0.0352)     & 0.6985(0.0393)   &FPCA-Kmeans         & 0.7537(0.0730)    & 0.6251(0.1027)     \\
			iterSubspace        & 0.8276(0.0364)     & 0.7286(0.0378)   &FPCA-GMM            & 0.8329(0.0874)    & 0.7343(0.1374)     \\\bottomrule
		\end{tabular}
	\end{center}
	\vskip -0.5cm
\end{table}

Third, as FMIclust and FSMIclust can be classified as unsupervised classification methods, it is valuable to compare their performance with supervised classification methods. According to \cite{zhang2019naive}, the simulation data were generated based on the model:
\begin{equation}
x_{i l}^{(y)}=\mu^{(y)}\left(t_{i l}\right)+\sum_{j=1}^J \xi_{i j}^{(y)} \phi_j\left(t_{i l}\right)+\epsilon_{i l}^{(y)}
\end{equation}
where $i=1, \ldots, n_y$ and $n_y$ is the number of samples for each class,  $y=1, \ldots, C$. We set $J=10$ for all simulations.  $\xi_{i j}^{(y)} \stackrel{i.i.d}{\sim} N(0, \lambda_j^{(y)})$  and $\epsilon_{i l}^{(y)}\stackrel{i.i.d}{\sim} N\left(0, 0.1\right)$. $\phi_1(t)=1, \phi_{2 r}(t)=\sqrt{2} \cos (2 r \pi t), \phi_{2 r+1}(t)=\sqrt{2} \sin (2 r \pi t)$, for $r=1,2, \ldots$. The 100 observation points of each curve are distributed at equal intervals [0,1]. The combination of the number of classes $C$, the mean function $\mu^{(y)}\left(t_{i l}\right)$ and eigenvalues $\lambda_j^{(y)}$ between classes are listed below:
\begin{itemize}
\item{Simulation 3: $C=2$, $\mu^{(1)}\left(t_{i l}\right)=\mu^{(2)}\left(t_{i l}\right)=0$; $\lambda_j^{(1)}=j$ and $\lambda_j^{(2)}=0.5j$;}
\item{Simulation 4: $C=2$, $\mu^{(1)}\left(t_{i l}\right)=0.3,\mu^{(2)}\left(t_{i l}\right)=0.6$; $\lambda_j^{(1)}=\lambda_j^{(2)}=j$;}
\item{Simulation 5: $C=3$, $\mu^{(y)}\left(t_{i l}\right)=0$; $\lambda_j^{(y)}=0.25yj$, $y=1,2,3$;}
\item{Simulation 6: $C=3$, $\mu^{(y)}\left(t_{i l}\right)=0.2y$; $\lambda_j^{(y)}=0.25yj$, $y=1,2,3$.}
\end{itemize}
For each simulation, we generated sample sizes of 300 and 600, with 60\% of the samples allocated as the training set and 40\% as the test set. Each simulated trajectory has a probability of $1/C$ to belong to group $y$. To evaluate the classification accuracy, we compare the proposed method with the Functional Naive Bayes model (FNB) \cite{zhang2019naive} and the functional logistic regression model (Flogstic) \cite{araki2009functional}. The regularization parameters in Flogstic are determined by the Bayesian information criterion (GBIC), while the kernel parameter in FNB is determined using 5-fold cross-validation. The experiment is repeated 100 times. In Table \ref{mv3}, we report the mean and standard deviation of classification accuracy rates.  FNB and Flogstic have a better classification effect, but the gap between FMIclust or FSMIclust and these supervised classification methods is not incomparable. Since FMIclust and FSMIclust do not use any class label information, we consider that these two methods have a fair efficiency of classification performance.

\begin{table}[ht]
	\begin{center}
		\caption{Classification accuracy rates with different methods }
		\vskip -0.5cm
		\footnotesize
		\begin{tabular}{cccccccccc}\label{mv3}
			\toprule
			Method    & Case                   & Simulation 3   & Simulation 4   & Simulation 5   & Simulation 6   \\\midrule
			\multirow{2}{*}{FNB}       & N=300 & 0.9332(0.0334) & 0.8791(0.0278) & 0.8431(0.0376) & 0.8692(0.0470) \\
			& N=600 & 0.9471(0.0291) & 0.8979(0.0251) & 0.8566(0.0269) & 0.8927(0.0381) \\
			\multirow{2}{*}{Flogstic}  & N=300 & 0.9126(0.0193) & 0.8401(0.0272) & 0.8492(0.0241) & 0.8510(0.0288) \\
			& N=600 & 0.9310(0.0162) & 0.8664(0.0249) & 0.8622(0.0192) & 0.8747(0.0255) \\
			
\multirow{2}{*}{FMIclust}  & N=300 & 0.8498(0.0294) & 0.8053(0.0341) & 0.7548(0.0259) & 0.7949(0.0325) \\
			               & N=600 & 0.8517(0.0218) & 0.8329(0.0273) & 0.7669(0.0385) & 0.8137(0.0271) \\
\multirow{2}{*}{FSMIclust} & N=300 & 0.8345(0.0267) & 0.7862(0.0391) & 0.7527(0.0282) & 0.8043(0.0387) \\
			               & N=600 & 0.8464(0.0218) & 0.8112(0.0252) & 0.7671(0.0311) & 0.8244(0.0283)\\\midrule
		\end{tabular}
	\end{center}
	\vskip -0.7cm
\end{table}

In order to assess the effectiveness of the proposed methods for estimating the number of clusters (C), we conducted a further analysis on the six simulations mentioned earlier. Each simulation was repeated 100 times and the model selection procedure was implemented over the 100 replications. Table \ref{mc52} summarizes the rates of underestimation (U), success (S), and overestimation (O), as well as the mean and standard deviation (std) of the estimated number of clusters on the simulated data sets. The results demonstrate that the proposed method is capable of properly
determining the appropriate number of clusters.

\begin{table}[ht]
	\footnotesize
	\begin{center}
		\caption{Rates of underestimating (U), success (S), and overestimating (O) by clustering number selection criteria on the simulation data sets in 100 replications}
		\vskip -0.5cm
		\footnotesize
		\begin{tabular}{ccccccccc}\label{mc52}
			\toprule
			\multirow{2}{*}{Case} & \multicolumn{4}{c}{FMIclust} & \multicolumn{4}{c}{FSMIclust} \\
			& U  & S   & O   & mean (std)  & U  & S   & O   & mean(std)    \\\midrule
			Simulation 1(C=2)     & 0  & 88  & 12  & 2.15 (0.43) & 0  & 84  & 16  & 2.22 (0.54)  \\
			Simulation 2(C=3)     & 4  & 82  & 14  & 3.12 (0.47) & 2  & 79  & 19  & 3.17 (0.43)  \\
			Simulation 3(C=2)     & 0  & 85  & 15  & 2.19 (0.48) & 0  & 86  & 14  & 2.14 (0.48)  \\
			Simulation 4(C=2)     & 0  & 81  & 19  & 2.24 (0.53) & 0  & 77  & 23  & 2.30 (0.61)  \\
			Simulation 5(C=3)     & 3  & 79  & 18  & 3.20 (0.57) & 4  & 74  & 22  & 3.21 (0.55)  \\
			Simulation 6(C=3)     & 6  & 74  & 20  & 3.16 (0.54) & 7  & 69  & 24  & 3.19 (0.58) \\\midrule
		\end{tabular}
	\end{center}
	\vskip -0.5cm
\end{table}

Finally, in order to investigate the time-costs of the proposed method, we revisit Simulation 1 and Simulation 2. The computation time of FMIclust and FSMIclust corresponds to the time required to compute a clustering solution after model selection has been completed. The average CPU computation times of 100 runs are presented in Table \ref{mc53}, with units of measurement in seconds (secs) and minutes (mins). The results demonstrate that the two proposed methods are competitive when compared with some of the existing functional data clustering methods.

\begin{table}[ht]
	\footnotesize
	\begin{center}
		\caption{ Computation time of FMIclust and FSMIclust}
		\vskip -0.5cm
		\footnotesize
	\begin{tabular}{ccccccccc}\label{mc53}
	\toprule
	Methods      & Simulation 1   & Simulation 2   & Methods   & Simulation 1   & Simulation 2   \\\midrule
	FMIclust     & 0.8592819 secs & 1.3294320 secs & funclust  & 15.003090 secs & 1.1034760 mins \\
	FSMIclust    & 0.4656234 secs & 0.6328394 secs & funHDDC   & 0.6750782 secs & 0.7959390 secs \\
	iterSubspace & 0.2134280 secs & 0.2712359 secs & waveclust & 10.593950 secs & 33.527770 secs \\
	fscm         & 4.0594690 secs & 8.6684790 secs & distclust & 1.1341190 mins & 2.5935510 mins \\
	B-Kmeans     & 0.1927519 secs & 0.2457172 secs & B-GMM     & 0.2278698 secs & 0.3180500 secs \\
	FPCA-Kmeans  & 0.1229380 secs & 0.1849319 secs & FPCA-GMM  & 0.1528579 secs & 0.1828519 secs \\\midrule
\end{tabular}
	\end{center}
	\vskip -0.5cm
\end{table}

\subsection{Real data analysis}
\subsubsection{The growth curves}
\begin{figure}[ht]
    \centering
    \subfigure[the growth curve (smooth)]{
		\begin{minipage}{0.42\textwidth}
            \centering
			\includegraphics[scale=0.26]{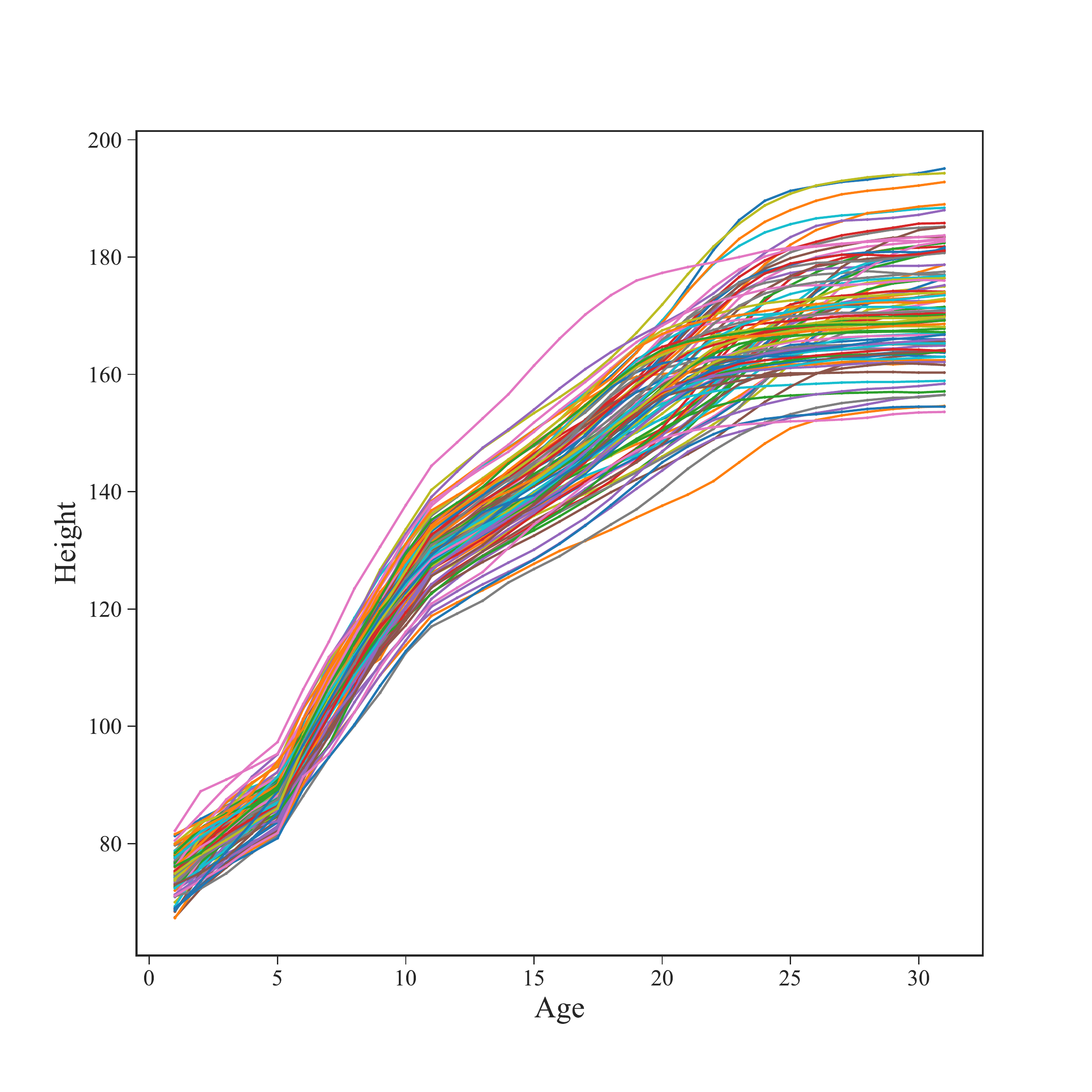}\label{mv51}
		\end{minipage}
    }
    \subfigure[the clustering result based on FMIclust]{
		\begin{minipage}{0.42\textwidth}
            \centering
			\includegraphics[scale=0.26]{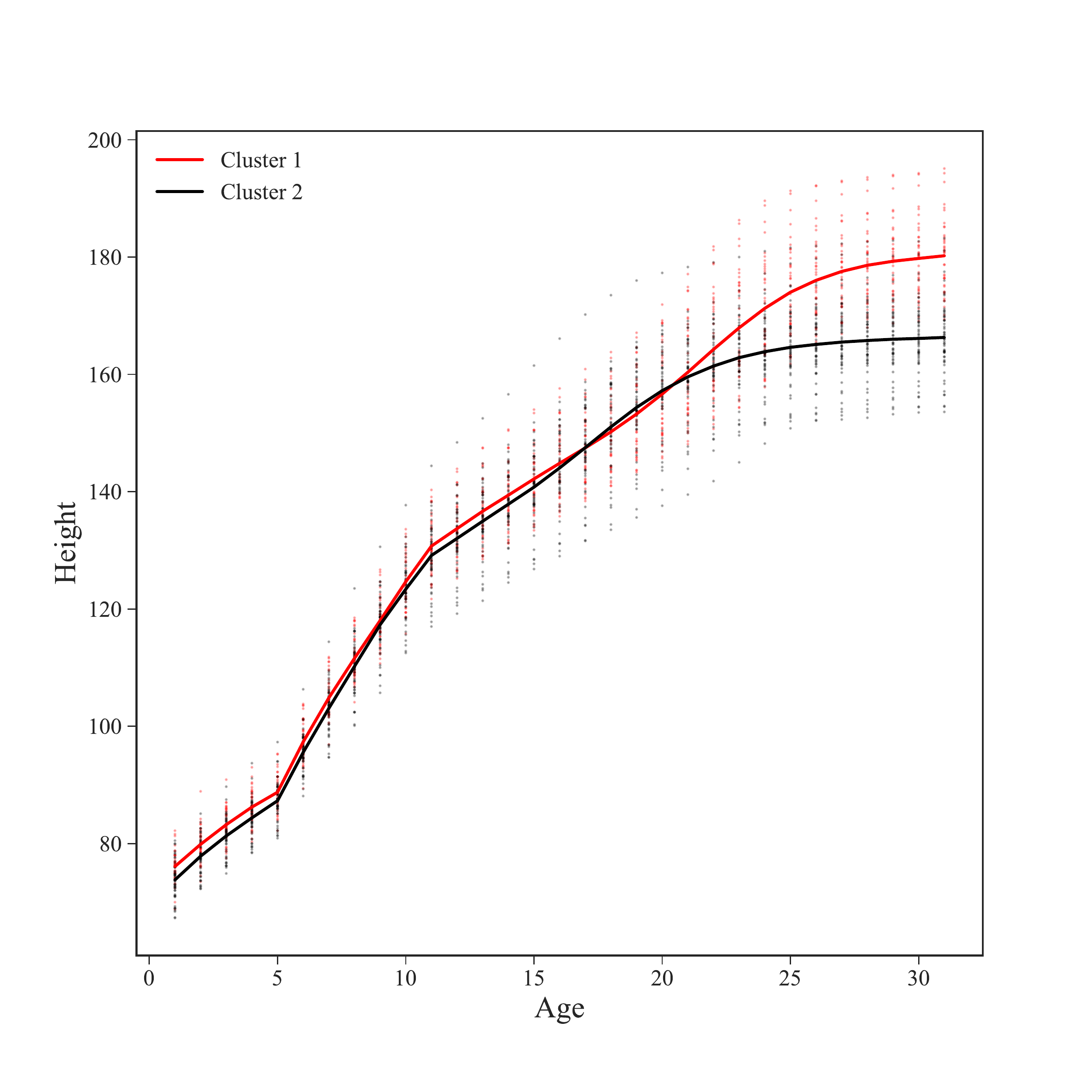}\label{mv52}
		\end{minipage}
   }
	\centering
\caption{Growth curve analysis}
\end{figure}

The Growth curve dataset (Growth) is derived from the renowned Berkeley Growth Research study \cite{tuddenham1954physical} and is currently accessible in the ``fda'' package of R.  In this dataset, the heights of 54 girls and 39 boys were measured at 31 stages, from 1 to 18 years. See Figure.\ref{mv51} for more details. The primary objective is to cluster the growth curves and subsequently ascertain whether the resultant clusters exhibit discernible gender disparities \cite{jacques2013funclust}.

\begin{table}[ht]
\footnotesize
\begin{center}
\caption{PF and ARI under different methods (Growth Data) }
\vskip -0.5cm
\footnotesize
\begin{tabular}{cccccccccc}\label{mv4}
\toprule
method  & FMIclust & FSMIclust & funclust   & funHDDC & waveclust     & iterSubspace \\\midrule
PF    & 0.9698   & 0.9587    & 0.6988       & 0.9670    & 0.8544      & 0.9355    \\
ARI   & 0.7951   & 0.7601    & 0.1876       & 0.7907    & 0.6302      & 0.7560    \\\midrule
method  & fscm   & distclust & B-Kmeans     & B-GMM     &FPCA-Kmeans  & FPCA-GMM  \\\midrule
PF    & 0.9227   & 0.8467    & 0.7845       & 0.7549    & 0.6443      & 0.9518    \\
ARI   & 0.7146   & 0.6073    & 0.4560       & 0.4274    & 0.1471      & 0.7586    \\\midrule
\end{tabular}
\end{center}
\vskip -0.6cm
\end{table}

On the one hand, the clustering results of each method are presented in Table \ref{mv4}. Notably, both FMIclust and FSMIclust exhibited exceptional performance, with FMIclust achieving the highest ARI and PF scores. On the other hand, when gender information was unknown (but the existence of two clusters was known), the average growth curves of the two clusters were computed using the class labels generated by the FMIclust method with the best performance, as depicted in Figure.\ref{mv52}. Over the entire growth period, the growth pattern of cluster 1 and cluster 2 was found to differ.  Specifically, in the early growth stage, the average growth curves of the two types maintained a nearly parallel upward trend with minor differences, indicating that the growth rates of the two groups during the early growth stage were comparable.  From 5 to 11 years of age, the change patterns of the average growth curves of the two clusters differed, with Cluster 1 exhibiting a higher growth curve than Cluster 2.  After 13 years of age, the growth rate of cluster 2 gradually decreased, resulting in an overall stabilization of growth.  In contrast, cluster 1 displayed a continued growth pattern during this period.  Given the well-known fact that men are generally taller than women, we suggest that cluster 1 represents males, and cluster 2 represents females. In summary, the proposed method can effectively distinguish the differences in growth patterns between men and women from growth curves. These differences may be attributed to a variety of factors, such as the varying hormone secretion patterns between men and women.

\subsubsection{ECG data}
\begin{figure}[ht]
\centering
\subfigure[Electrocardiogram curve (smooth)]{
\begin{minipage}[b]{0.42\linewidth}
\centering
\includegraphics[scale=0.26]{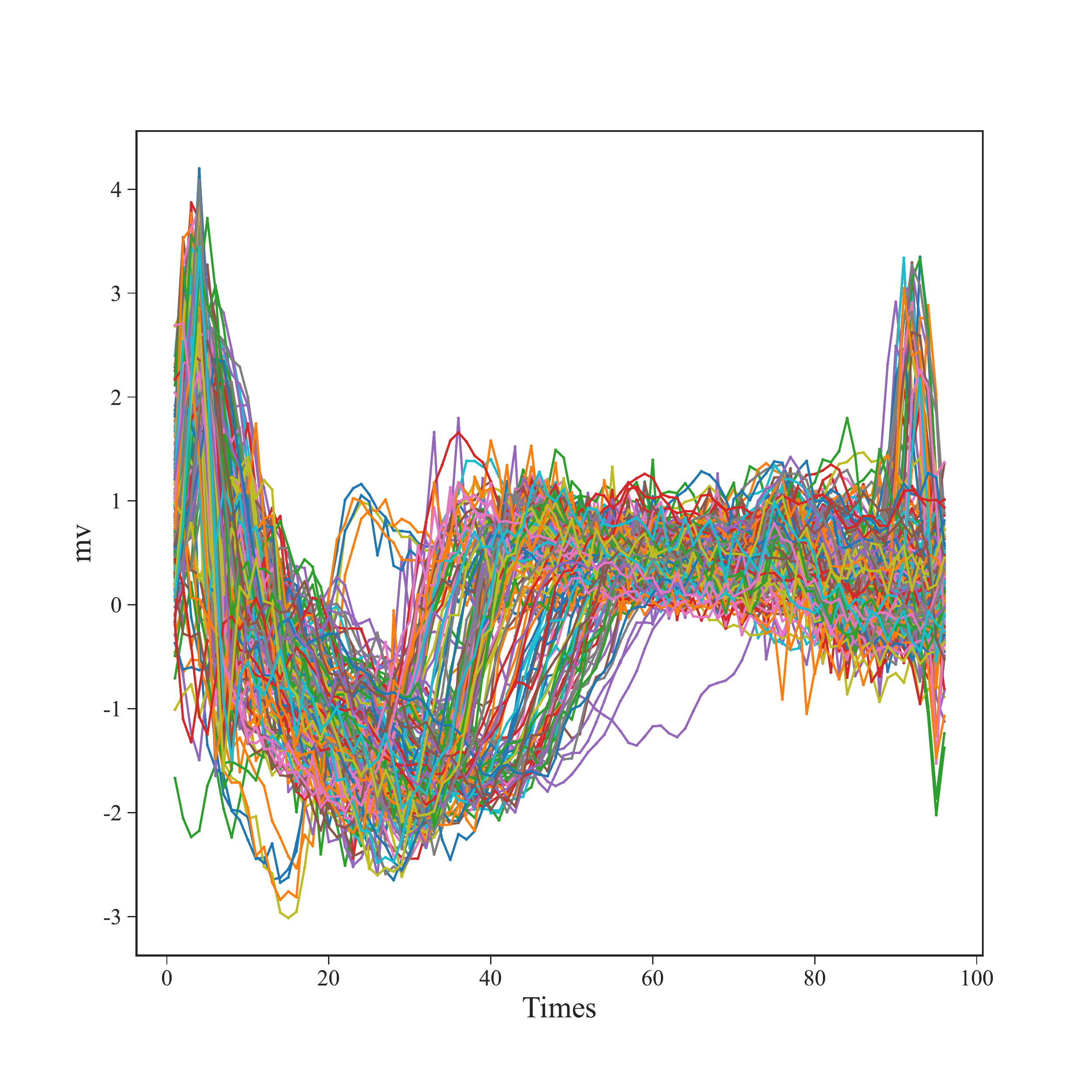}\label{mv71}
\end{minipage}
}
\subfigure[the clustering result based on FMIclust]{
\begin{minipage}[b]{0.42\linewidth}
\centering
\includegraphics[scale=0.26]{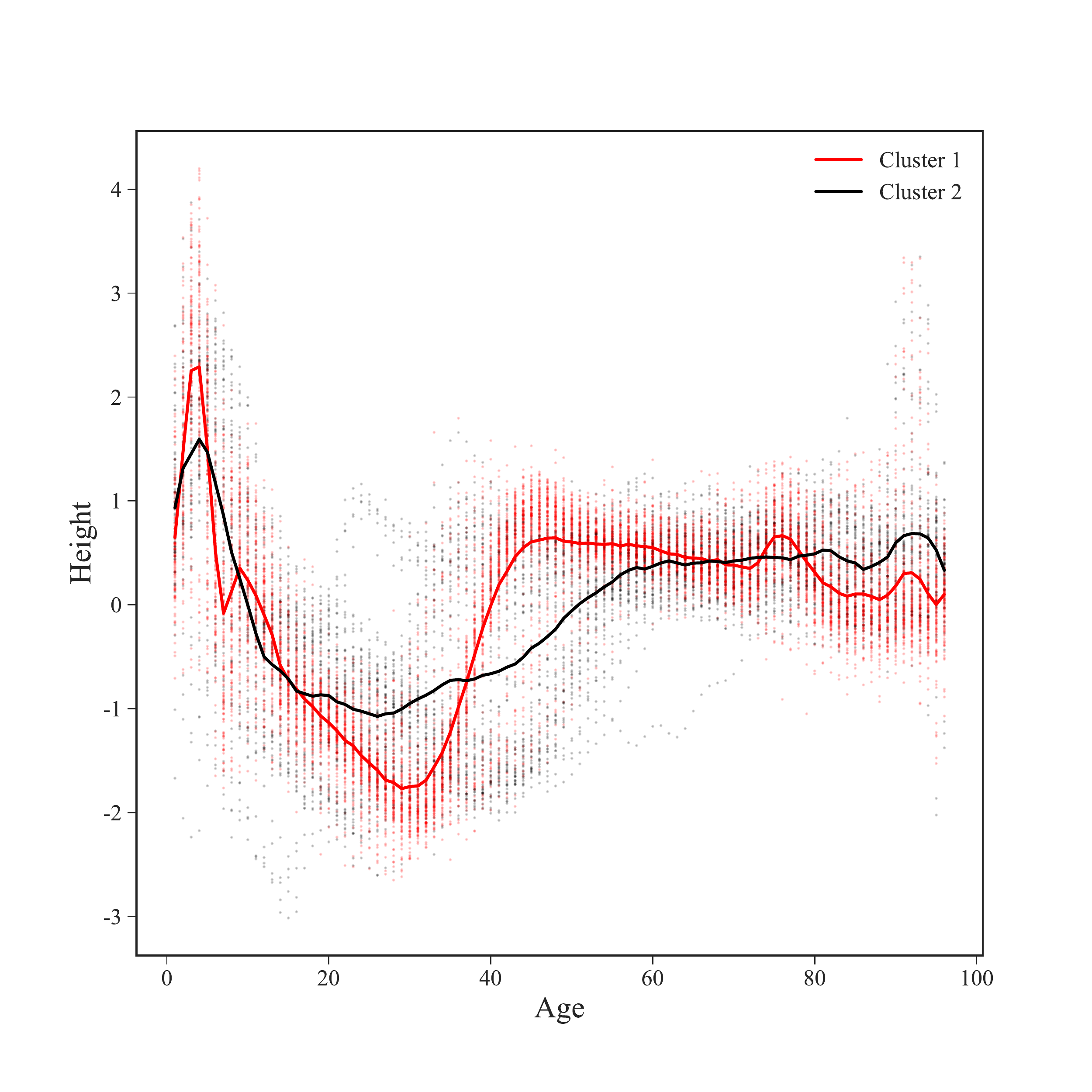}\label{mv72}
\end{minipage}
}
\caption{ECG curve analysis}
\end{figure}

The ECG dataset (ECG) used in this study was obtained from the UCR Time Series Classification and Clustering website (\url{https://www.cs.ucr.edu/~eamonn/time_series_data_2018/}).  The dataset comprises 200 electrocardiogram recordings from two groups of patients, as shown in Figure.\ref{mv71}. Among the tracks were patients with myocardial infarction and subjects with a normal heartbeat. The objective of this study is to cluster the ECG curves, assess the consistency of the clustering results with the true labels, and analyze the differences between myocardial infarction and normal heartbeat based on the clustering results \cite{bouveyron2011model}.

\begin{table}[ht]
\footnotesize
\begin{center}
\caption{PF and ARI under different methods (ECG Data)}
\vskip -0.5cm
\begin{tabular}{cccccccccc}\label{mv6}
\toprule
method  & FMIclust & FSMIclust & funclust     & funHDDC   & waveclust   & iterSubspace \\\midrule
PF    & 0.8751   & 0.8530    & 0.8418       & 0.7542    & 0.7644      & 0.8355       \\
ARI   & 0.7251   & 0.7001    & 0.6776       & 0.6107    & 0.6202      & 0.6960       \\\midrule
method  & fscm     & distclust & B-Kmeans     & B-GMM     &FPCA-Kmeans  & FPCA-GMM      \\\midrule
PF    & 0.8227   & 0.7467    & 0.7445       & 0.8150    & 0.7673      & 0.8118        \\
ARI   & 0.6846   & 0.6073    & 0.6060       & 0.6774    & 0.6471      & 0.6686        \\\bottomrule
\end{tabular}
\end{center}
\vskip -0.6cm
\end{table}

The clustering results of each method are presented in Table \ref{mv6}. Notably, both FMIclust and FSMIclust exhibited exceptional performance, with FMIclust achieving the highest ARI and PF scores. Furthermore, assuming that the true labels are unknown and based on professional medical knowledge, the ECG tracks of 200 subjects were defined into myocardial infarction and normal
heartbeat based on the class label generated by FMIclust. As shown in Figure.\ref{mv72}, the ECG recording marked in red exhibits abnormal negative waves, while the one marked in black is relatively normal. This abnormality, known as the ``diagnostic Q wave" in medical diagnosis, is a key diagnostic feature with significant value. Myocardial infarction can be further classified into various types based on the ``diagnostic Q wave" including inferior myocardial infarction, anterior inter-wall myocardial infarction, anterior wall myocardial infarction, extensive anterior wall myocardial infarction, lateral wall myocardial infarction, positive posterior wall myocardial infarction, and right ventricular myocardial infarction. Therefore, the FMIclust method can diagnose myocardial infarction by identifying characteristic changes in the ECG recording, without relying on clinical symptoms or elevated markers of myocardial necrosis. 

\subsubsection{The Canadian temperature data}
\begin{figure}[ht]
\centering
\subfigure[Daily temperature curve (smooth)]{
\begin{minipage}[b]{0.42\linewidth}
\centering
\includegraphics[scale=0.26]{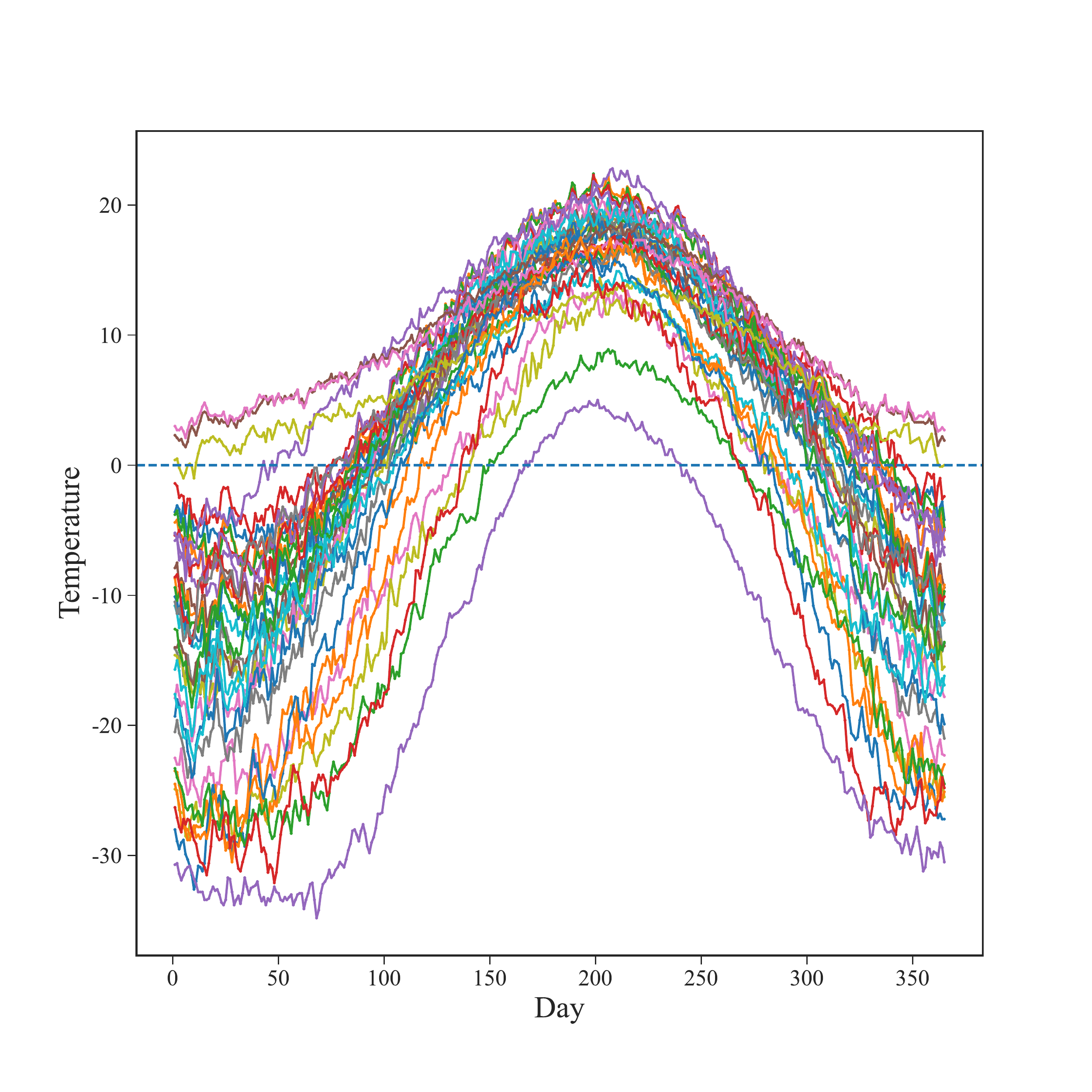}\label{mv91}
\end{minipage}
}
\subfigure[Clustering result based on FSMIclust]{
\begin{minipage}[b]{0.42\linewidth}
\centering
\includegraphics[scale=0.26]{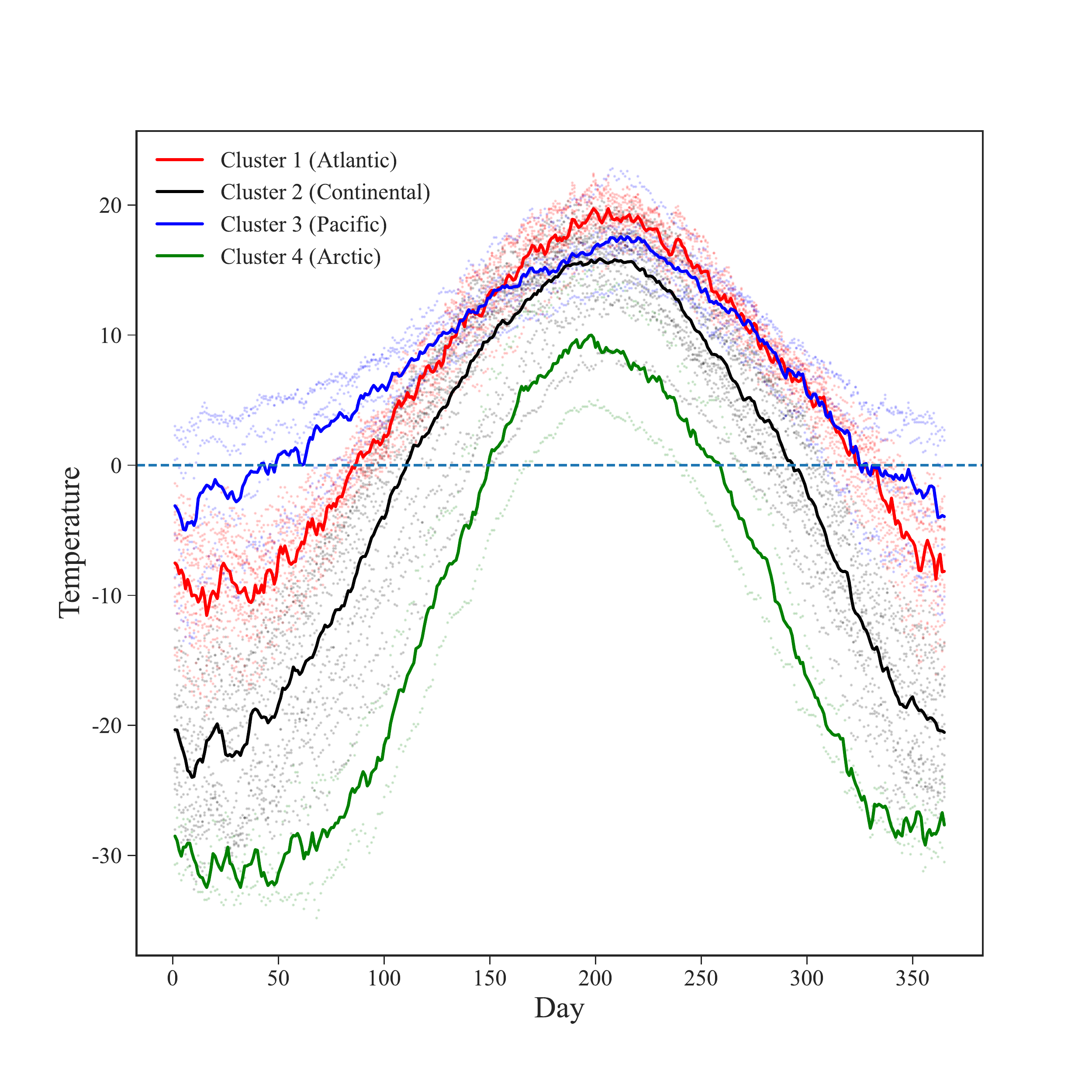}\label{mv92}
\end{minipage}
}
\caption{Analysis of Canadian temperature data}
\end{figure}

The Canadian Weather dataset records daily average temperature curves for 35 Canadian cities. Specific data can be obtained from the R package called ``fda''. The annual daily mean temperature distribution of these cities is shown in Figure.\ref{mv91}. The aim of this study is to cluster these cities into four climatic zones (Atlantic, Arctic, Pacific, and continental) based on their daily average temperature curves. Subsequently, the geographical interpretation of the clustering results will be analyzed \cite{centofanti2022smooth, wang2016functional}. 

\begin{table}[ht]
\begin{center}
\footnotesize
\caption{PF and ARI under different methods (Canadian temperature Data)}
\vskip -0.4cm
\begin{tabular}{cccccccccc}\label{mv8}
\toprule
method  & FMIclust & FSMIclust & funclust     & funHDDC   & waveclust   & iterSubspace \\\midrule
PF    & 0.8857   & 0.9143    & 0.8571       & 0.8571    & 0.8286      & 0.8857       \\
ARI   & 0.6984   & 0.7840    & 0.6774       & 0.6528    & 0.5680      & 0.7414       \\\midrule
method  & fscm     & distclust & B-Kmeans     & B-GMM     &FPCA-Kmeans  & FPCA-GMM      \\\midrule
PF    & 0.8571   & 0.8286    & 0.7714       & 0.8286    & 0.7714    & 0.8286        \\
ARI   & 0.7169   & 0.6246    & 0.4680       & 0.6785    & 0.5083    & 0.5830        \\\bottomrule
\end{tabular}
\end{center}
\vskip -0.4cm
\end{table}

\begin{figure}[ht]
    \centering
    \includegraphics[scale=0.3]{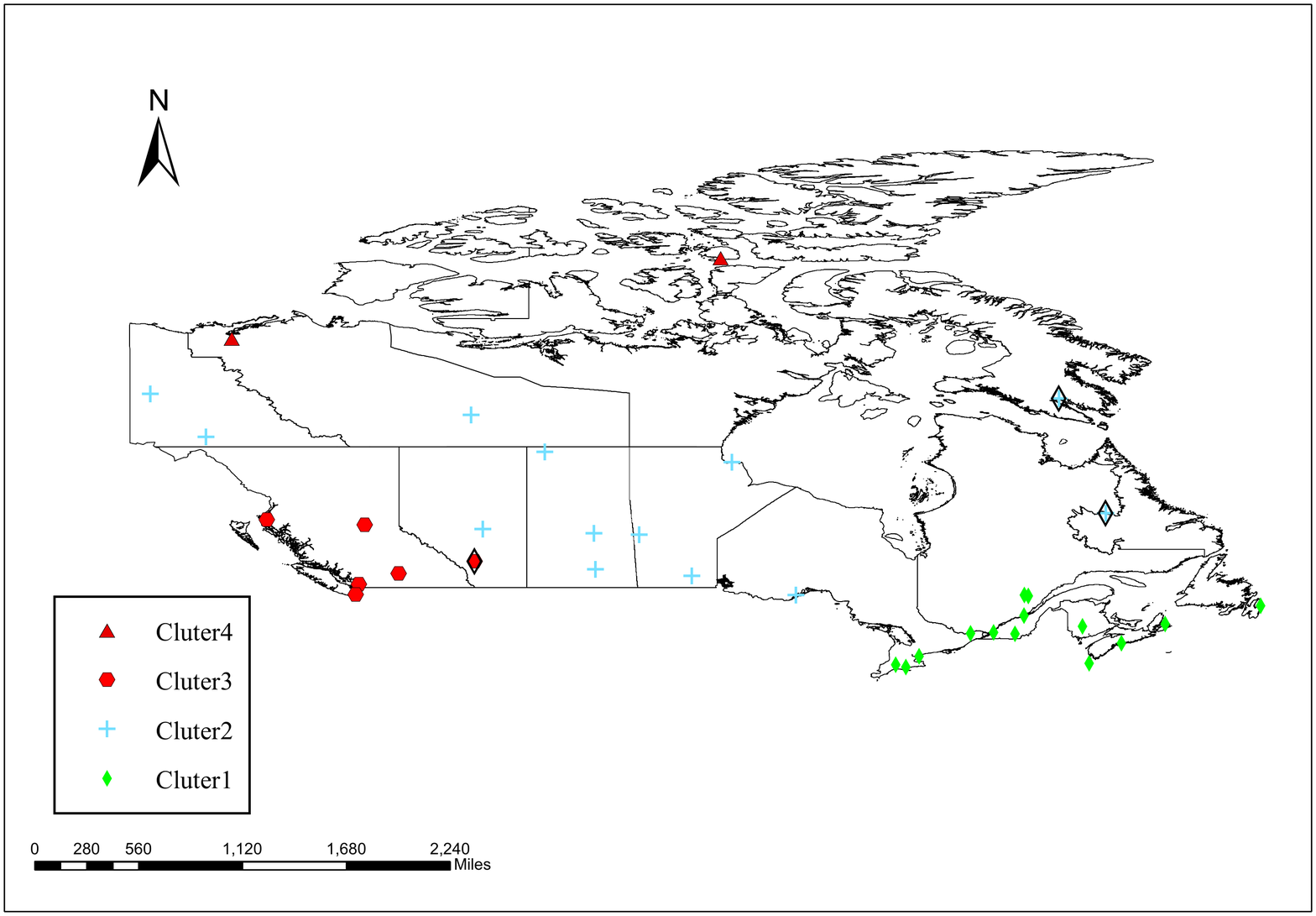}
    \caption{Geographical distribution based on FSMIclust}
\label{mv10}
\end{figure}

On the one hand, the clustering results of each method are shown in Table \ref{mv8}. Specifically, Both FMIclust and FSMIclust showed excellent performance, among which PF
and ARI were the highest in FSMIclust. On the other hand, 35 cities in Canada are divided into four distinct climatic zones using the FSMIclust method. The geographical distribution characteristics of these four zones are displayed in Figure.\ref{mv10}, with black diamonds representing misclassified cities. For now, let us disregard the three misclassified cities (Schefferville, Calgary, and Iqaluit). Firstly, the cities denoted by green diamonds (cluster 1) are located on the Atlantic coast and belong to the Atlantic climate zone (maritime climate). The average daily temperature in summer is relatively high, and the annual temperature ranges from $-10^{\circ}C$ to $25 ^{\circ}C$, which is consistent with the daily average curve characteristics (red marks) shown in Figure.\ref{mv92}. The cities marked with red dots (cluster 2) are located on the Pacific coast, and they belong to the Pacific climate zone (temperate maritime climate). The temperature remains above zero for most of the year, and the average annual temperature is relatively high compared to other climatic zones, with a relatively slow annual temperature change. This is in line with the daily average curve (blue marks) features shown in Figure.\ref{mv92}. Cities marked with blue plus signs (cluster 3) are located in the interior of Canada and belong to the continental climate zone (continental climate). Under continental climate conditions, the average annual temperature varies greatly, with rapid warming from spring to summer and rapid cooling from summer to winter, which is contrary to the Atlantic and Pacific climatic zones, characterized by marine climates. The marine climate is typified by slow warming from spring to summer and slow cooling from summer to winter. These characteristics are also readily discernible from the daily mean temperature curve (black marks) shown in Figure.\ref{mv92}. Finally, the cities marked with red triangles (cluster 4) are situated in the Arctic region at high latitudes. The winters are long and cold, while summers are brief and warm, with temperatures remaining below $0^{\circ}C$ for most of the year. This corresponds to the daily average curve characteristics (green marks) displayed in Figure.\ref{mv92}.

Out of the 35 cities analyzed in this study, three were misclassified by the proposed method. Specifically, Iqaluit, located in the Arctic, was categorized as being in the continental climate zone, while Calgary, situated in the continental climate zone, was classified as being in the Pacific climate zone. Similarly, Schefferville, located in the Atlantic climate zone, was categorized as being in the continental climate zone. These misclassifications can be attributed to the fact that these cities are located in transition zones between different climatic zones, making it challenging to differentiate their spatial characteristics. Despite these misclassifications, our proposed method outperformed other methods in terms of overall accuracy. Therefore, we contend that our proposed method is a reasonable and effective approach for classifying the climatic zones of the 35 cities in Canada.

\subsubsection{COVID-19 data in China}
\begin{figure}[ht]
    \centering
    \includegraphics[scale=0.3]{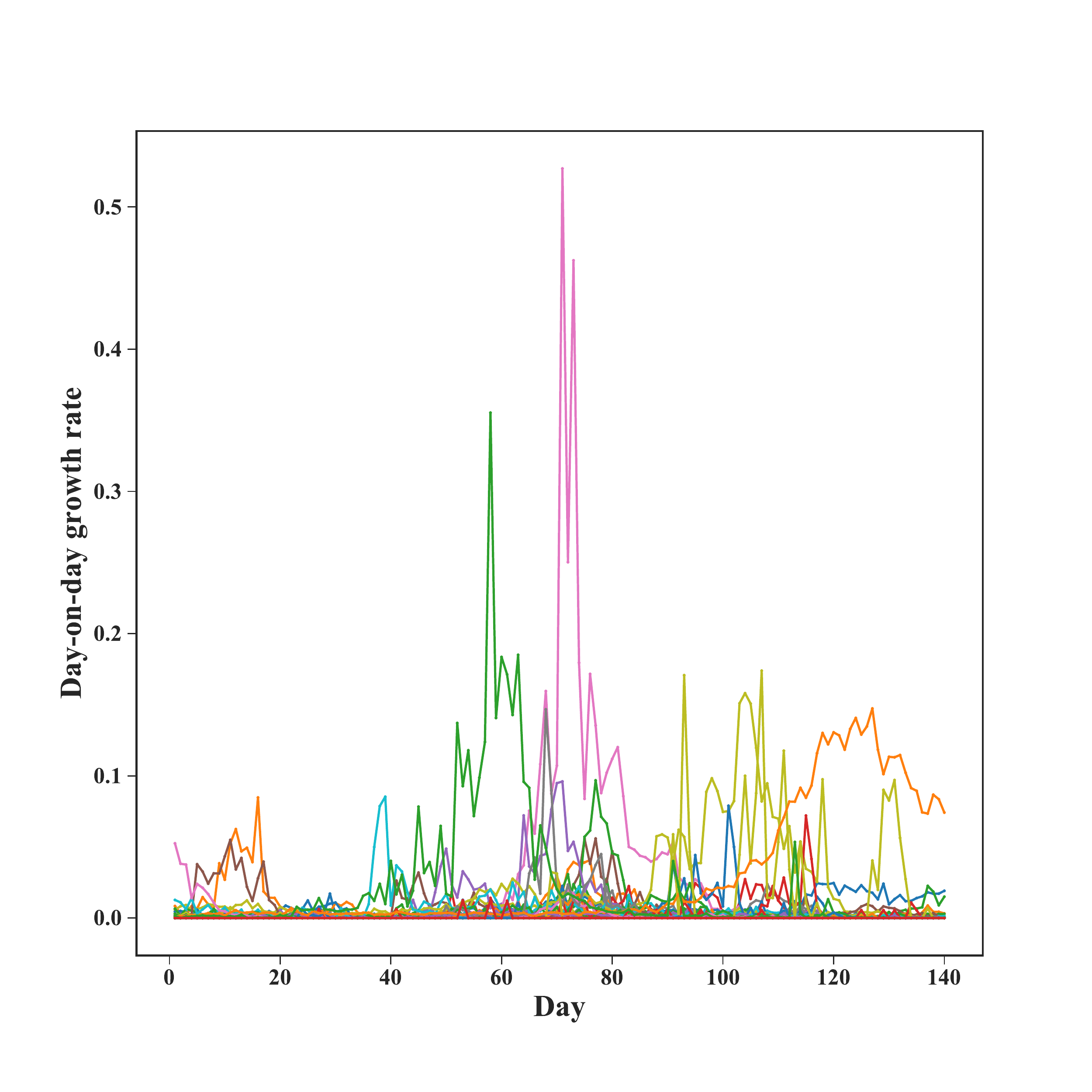}
\vskip -0.8cm
    \caption{the day-on-day growth rate of COVID-19 in China}
\label{vvv1}

\end{figure}

The COVID-19 dataset was collected from the daily COVID-19 notifications published by the National Health Commission of China and the health commission websites of 31 provinces (autonomous regions, municipalities, and corps) (\url{http://www.nhc.gov.cn}).  The data was acquired using crawler technology (\url{https://github.com/xinyuli11/Epidemic-data}). As the studies by Li \cite{li2020retrospective} and  Martin-Barreiro \cite{martin2021disjoint}, the day-on-day growth rate of the cumulative number of confirmed cases has been identified as an effective indicator for assessing the short-term trend of the epidemic.  Therefore, we have calculated the day-on-day growth rate of the cumulative number of confirmed cases in all provinces (autonomous regions, municipalities, and corps) in China from January 1st to May 20th, 2022 as shown in Figure.\ref{vvv1}. The aim of this study is to analyze the recent state of COVID-19 epidemic prevention and control across 31 provinces (autonomous regions, municipalities, and corps) in China, and propose recommendations for regional management based on the current status of epidemic prevention and control in each province.

\begin{figure}[ht]
\centering
\subfigure[spatial distribution of COVID-19 (FMIclust)]{
\begin{minipage}[b]{0.44\linewidth}
        \centering
\includegraphics[scale=0.22]{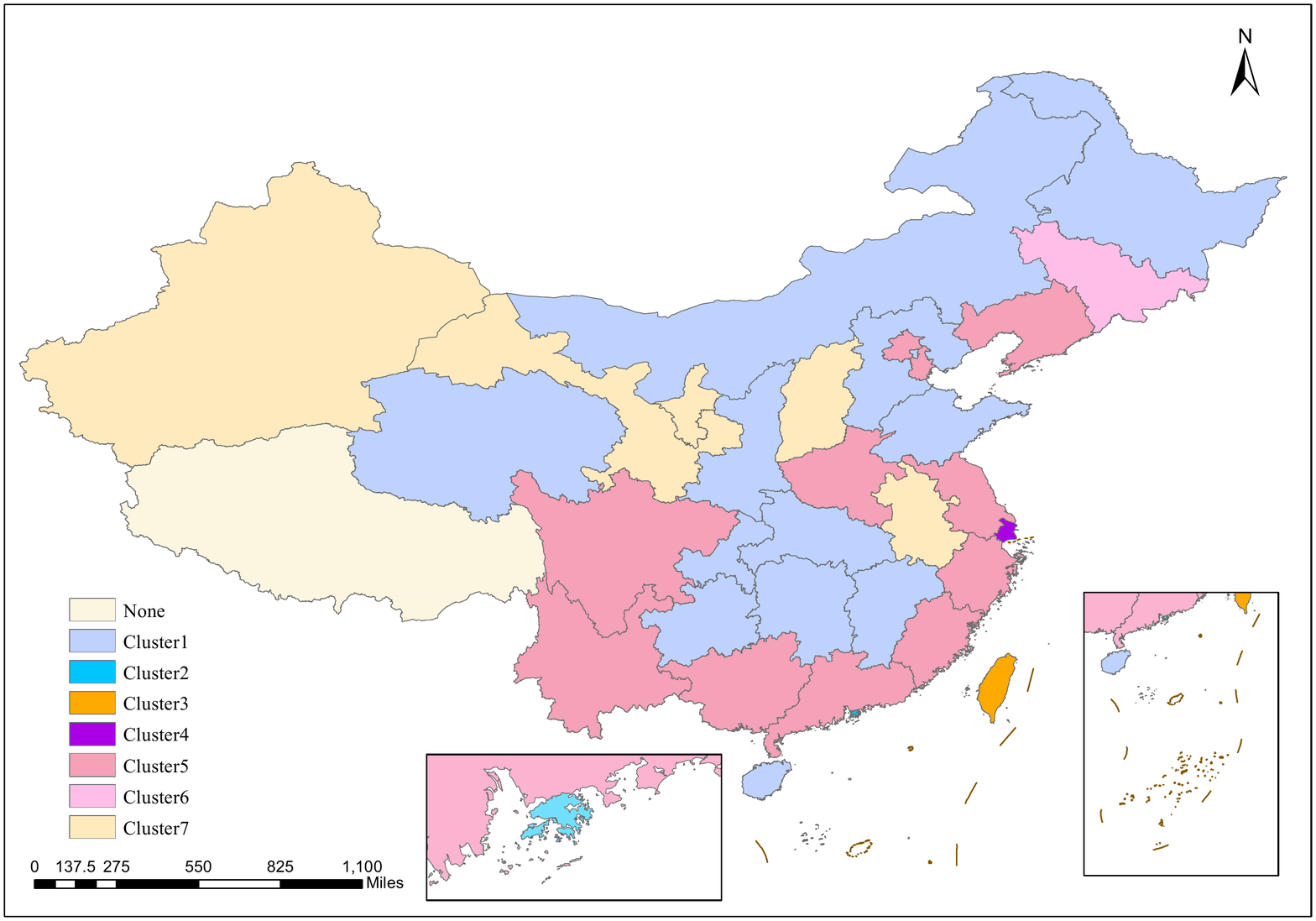}
\end{minipage}
}
\subfigure[spatial distribution of COVID-19 (FSMIclust)]{
\begin{minipage}[b]{0.44\linewidth}
        \centering
\includegraphics[scale=0.22]{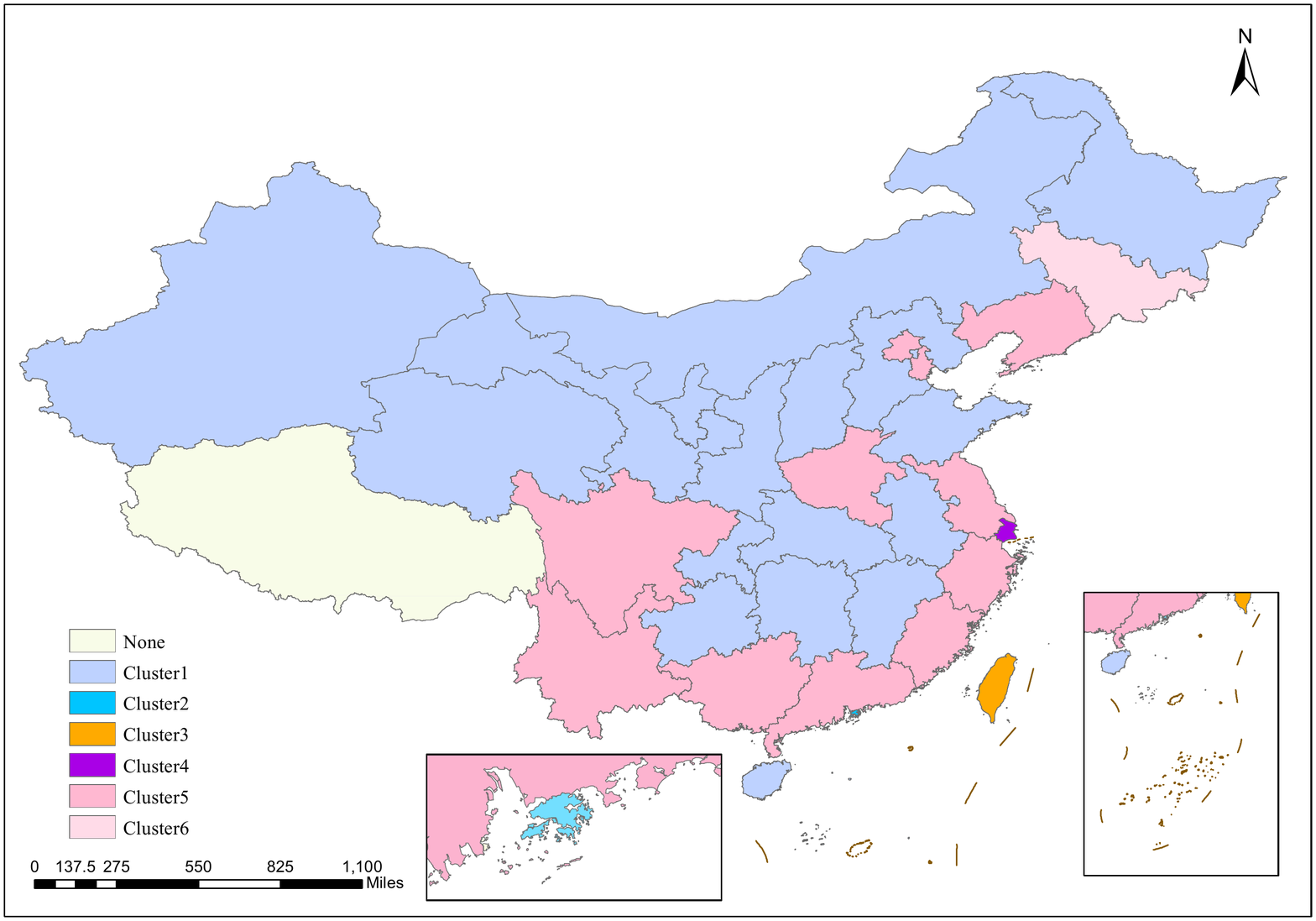}
\end{minipage}
}
\caption{spatial distribution of COVID-19 in China from 01.01 to 05.20,2022}
\label{vvv2}
\end{figure}

\begin{figure}[ht]
    \centering
    \includegraphics[scale=0.28]{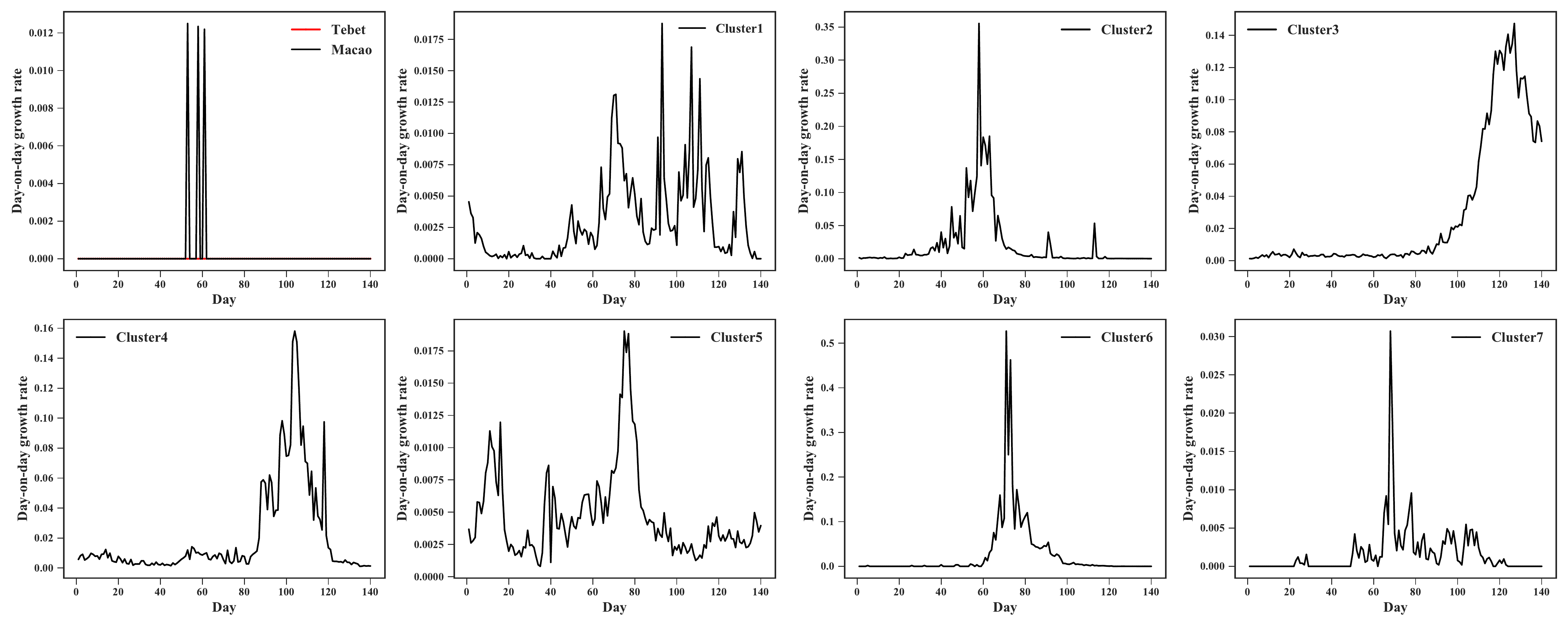}
    \caption{Average day-on-day growth rate (FMIclust)}
\label{vvv4}
\end{figure}

\begin{figure}[ht]
    \centering
    \includegraphics[scale=0.3]{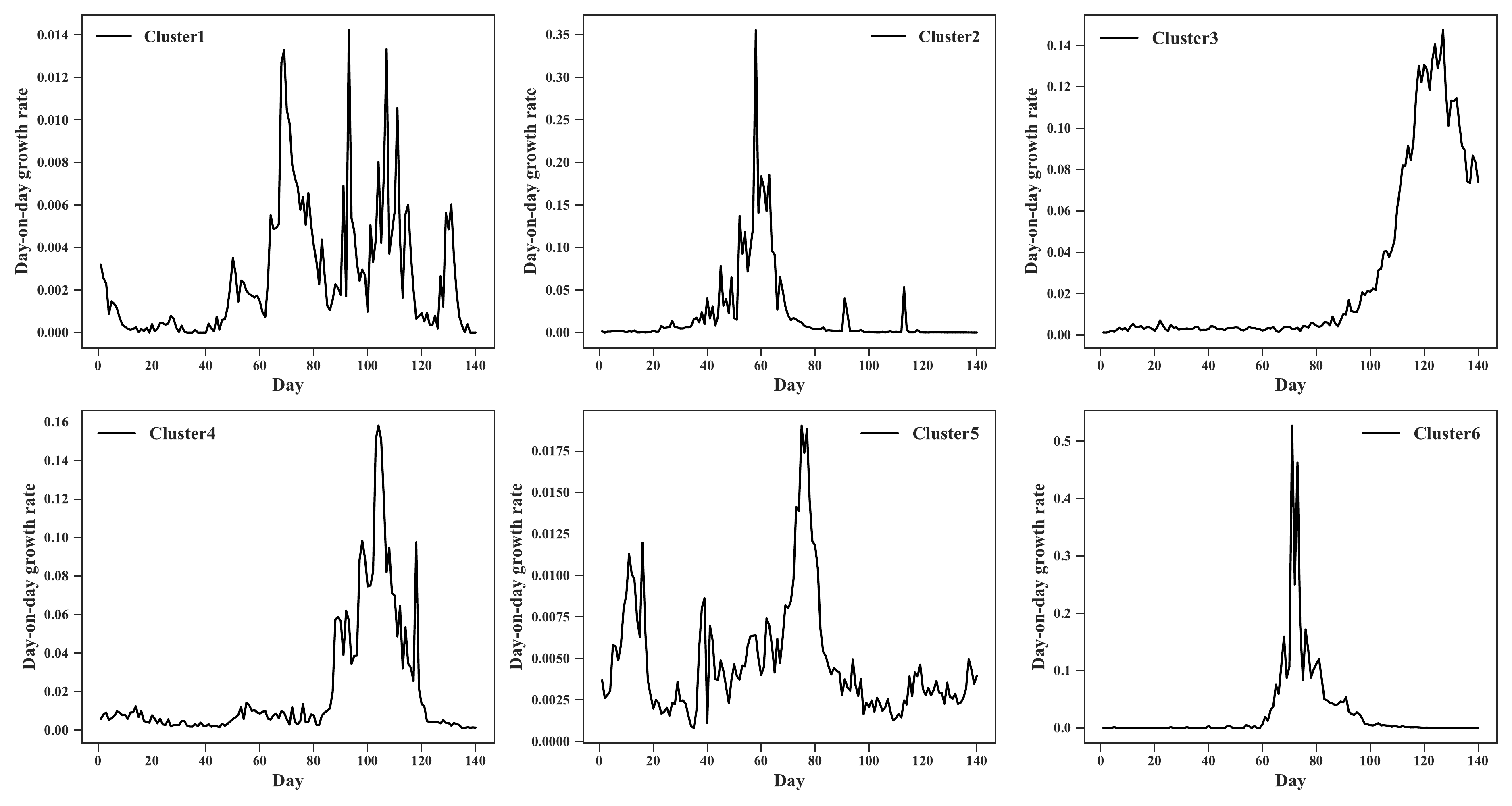}
    \caption{Average day-on-day growth rate (FSMIclust)}
\label{vvv5}
\end{figure}

From January 1 to May 20, 2022, Tibet, Ningxia, and Macao were excluded from the clustering analysis due to the daily growth rate of COVID-19 being almost zero. Based on the scree-test of Cattell, the optimal number of clusters for FMIclust was determined to be 7. The resulting clustering assignments are as follows: (1) Hebei, Heilongjiang, Jiangxi, Shandong, Hubei, Hunan, Hainan, Chongqing, Guizhou, Shaanxi, Qinghai, and Inner Mongolia;  (2) Hong Kong;  (3) Taiwan;  (4) Shanghai;  (5) Jilin;  (6) Beijing, Tianjin, Liaoning, Jiangsu, Zhejiang, Fujian, Henan, Guangdong, Guangxi, Sichuan, and Yunnan;  (7) Shanxi, Anhui, Gansu, and Xinjiang. Additionally, the number of clusters identified by FSMIclust was 6 based on the modified BIC criterion. Compared to the clustering results of FMIclust, FSMIclust combined cluster (1) and cluster (7), which correspond to low-risk areas with sporadic outbreaks and no serious outbreaks, respectively.  This combination is reasonable given the similarity of the two clusters.  The geographical distribution of the clusters is illustrated in Figure.\ref{vvv2}, while the average daily growth rates of COVID-19 in each cluster are presented in Figures.\ref{vvv4} and \ref{vvv5}.


Based on the clustering results obtained from FMIclust, cluster 1 exhibited a relatively low day-on-day growth rate but had a growth pattern characterized by frequent fluctuations with multiple growth peaks observed in March. The day-to-day growth rate of cluster 2 displayed a fluctuating upward trend starting from February, reaching its peak in early March, and then gradually declining. Cluster 3 demonstrated a sustained and rapid day-to-day growth rate from April, with a slight deceleration observed in May, but still maintaining a high level. Cluster 4's day-to-day growth rate persisted from March to April, peaked in April, but was soon brought under control. Cluster 5 experienced ongoing fluctuations in its day-to-day growth rate, with the most significant growth observed in March. Both cluster 6 and cluster 7 exhibited substantial increases in their day-to-day growth rates in March. However, the key difference between these two clusters is that cluster 6 was effectively contained in a short period, while cluster 7 still exhibited growth. Collectively, our findings suggest that the epidemic development in China between January 1 and May 20, 2022, exhibits diverse patterns of growth across different clusters.

In combination with the actual situation, it is evident that severe COVID-19 outbreaks occurred in Jilin Province, Hong Kong Special Administrative Region, Shanghai, and Taiwan Province from January 1 to May 20, 2022, and each region displayed unique epidemic development patterns. Except for Taiwan, although the growth rate of other regions was higher in the outbreak period and epidemic period, the descending trend was also fast in the later period. The growth rate tends to zero and the fluctuation is small, which indicates that epidemic prevention and control is efficient. The day-on-day growth rate of cluster 1 was relatively small, but it is worth noting that its fluctuation was relatively large, which may be because the cross-provincial transmission of the epidemic has not been completely cut off. The day-on-day growth rate and its fluctuation of cluster 7 were relatively small. This is partly due to China's population mobility pattern, and partly due to the scientific and precise epidemic prevention and control measures taken in these regions. In cluster 6, there were all outbreaks of different degrees in the period from March to April 2022. However, the day-on-day growth curve of these regions has the characteristics of relatively fast rising and falling speed, and when the growth rate drops to a low level, its fluctuation tends to 0. This may be related to the targeted measures taken by these regions to prevent imports from outside, prevent rebound from inside, and provide precise prevention and control and services. In addition, one of the reasons why the average day-on-day growth rate exceeded 0.5 is that the cumulative number of confirmed cases in these regions was relatively small before and then rose sharply after the outbreak.

In summary, the outbreak in China from January 01 to May 20, 2022, was characterized by several factors. Firstly, it persisted in a multifaceted situation, exhibiting both scale-up and sporadic clustered outbreaks. Secondly, it presented a complex situation of multi-source and multi-chain at multiple points. Despite these challenges, it is evident that China's provinces have implemented active and effective measures to control the outbreak, resulting in a gradual reduction of the disease risk in society. This illustrates that China has constructed a comprehensive and multi-level outbreak control system, which can achieve effective containment of the spread of the disease.

\section{Conclusions}\label{nn2}
In this paper, we proposed a new functional information-maximization clustering method that learns class-posterior probabilities in an unsupervised manner so that the mutual information (or squared loss mutual information) between data points and cluster assignments is maximized. A notable advantage of this approach is that classifier training is formulated as continuous optimization problems, which are substantially simpler than discrete optimization of cluster assignments. Moreover, we do not need to estimate the probability densities of Karhunen-Lo\`eve scores for different clusters and do not require the common eigenfunction assumption. The results of the simulation experiment and real data analysis show that the proposed method has universality and excellent performance. 

Next, we discuss the suitability of each of the two methods for different situations. Firstly, FMIclust is flexible in the choice of the conditional model $P(y|\boldsymbol{z}^{q}(x))$ and the regularization item $R(\lambda ;\boldsymbol{W})$. Compared to FSMIclust, FMIclust estimates the true class-prior probability in a data-driven fashion by iteratively performing clustering and updating the class-prior probabilities, making it a useful property in practice when the true class-prior probability is unknown. Secondly, the advantages of FSMIclust lie primarily in the solving efficiency of optimization problems. In particular, the sparse local-scaling kernel used in FSMIclust has proven to be useful in experiments, resulting in a sparse kernel matrix that allows for efficient computation of the clustering solution (i.e., solving a kernel eigenvalue problem). When the class prior information is sufficient, FSMIclust can be used to obtain clustering solutions with higher computational efficiency. Furthermore, since both proposed methods are developed based on information maximization clustering, we can comprehensively compare the clustering results of the two methods based on the actual situation to determine the final clustering assignments. Overall, the choice of method depends on the specific characteristics of the data and the goals of the analysis. FMIclust may be preferable when the true class-prior probability is unknown and when greater flexibility in modeling is required, while FSMIclust may be more appropriate when computational efficiency is a major concern and when class prior information is sufficient.

There are several important aspects for future research in the field of functional information maximization clustering. Firstly, with regard to the FMIclust method, it may be worthwhile to investigate the use of different combinations of regularization terms and conditional models to handle more complex data cases. For example, when the number of functional principal components is large, the use of a sparse term to penalize the logistic regression parameter, and the proximal gradient (PG) method to solve the sparse optimization. Alternatively, consideration may be given to assuming that the conditional model of the class-posterior probability is multivariate kernel logistic regression or other models \cite{krause2010discriminative}. Secondly, we found empirically that the active functions ``sigmoid (or softmax)" and ``Max" perform well and achieve similar experimental effects in the FSMIclust method. As an example, we reconsider Simulation 1 and Simulation 2. For Simulation 1, the PF and ARI of FSMIclust based on ``sigmoid'' is 0.9265(0.0337) and 0.8534(0.0276), which is no significant difference (1\%) with the results of FSMIclust based on ``Max'' (PF: 0.9318(0.0321); ARI: 0.8632(0.0231)). For Simulation 2, the PF and ARI of FSMIclust based on ``softmax'' is 0.8994(0.0224) and 0.7965(0.0306), which is also no significant difference (1\%) with the results of FSMIclust based on ``max''(PF: 0.8931(0.0237); ARI: 0.7902(0.0282)). Further investigation into the performance of different activation functions in future work may be beneficial. Thirdly, theoretically elucidating statistical consistency of the proposed method as well as investigating the perturbation stability in more details is also an important challenge. Fourthly, It is also worth discussing how to extend the functional information maximization clustering framework to the case of multivariate functional data by effectively fusing the clustering information  of different random function variables. Lastly, through functional cluster analysis of COVID-19 epidemic data, it has been noted that all provinces exhibit similar functional characteristics during different time periods, akin to gene expression clustering analysis in biostatistics. The bidirectional clustering method (Bicluster) may prove to be a valuable tool for investigating this similarity \cite{fang2022biclustering, mankad2014biclustering, ma2017concave}. Further research into the development of a bidirectional clustering analysis method for functional data within the information maximization framework is also an important topic to the field.


\bibliographystyle{tfnlm}
\bibliography{name}
\end{document}